\documentclass[fleqn,usenatbib]{mnras}

\usepackage{newtxtext,newtxmath}
\usepackage[T1]{fontenc}

\DeclareRobustCommand{\VAN}[3]{#2}
\let\VANthebibliography\thebibliography
\def\thebibliography{\DeclareRobustCommand{\VAN}[3]{##3}\VANthebibliography}

\usepackage{graphicx}	% Including figure files
\usepackage{amsmath}	% Advanced maths commands
\usepackage{hyperref}
\usepackage{cprotect}
\usepackage{multirow}

\title[On the road to the radius valley]{On the road to the radius valley: distinguishing between gas dwarfs and water worlds with young transiting exoplanets}

\author[J. G. Rogers]{
James G. Rogers$^{1}$\thanks{E-mail: jr2011@cam.ac.uk}
\\
$^{1}$Institute of Astronomy, University of Cambridge, Madingley Road, Cambridge CB3 0HA, United Kingdom
}

\date{Accepted XXX. Received YYY; in original form ZZZ}

\pubyear{\the\year{}}

\begin{document}
\label{firstpage}
\pagerange{\pageref{firstpage}--\pageref{lastpage}}
\maketitle

\begin{abstract}
The detection of young transiting exoplanets represents a new frontier in our understanding of planet formation and evolution. For the population of observed close-in sub-Neptunes, two proposed formation pathways can reproduce their observed masses and radii at $\sim$~Gyr ages: the ``gas dwarf'' hypothesis and the ``water world'' hypothesis. We show that a sub-Neptune's size at early ages $\lesssim 100$~Myrs is strongly dependent on the bulk mean molecular weight within its envelope. As a result, gas dwarfs and water worlds should diverge in size at early ages since the mean molecular weight of gas dwarf envelopes is predicted to be smaller than that of water worlds. We construct population models under both scenarios that reproduce \textit{Kepler} demographics in the age range $\sim1-10$~Gyrs. We find tentative evidence that the gas dwarf model is more consistent with the small population of young exoplanets $< 40$~Myrs from \textit{TESS}. We show that planet radius is relatively insensitive to planet mass for young, puffy sub-Neptunes, meaning that well-characterised masses are not necessarily required to exploit the effects of mean molecular weight at the population level. We confirm the predicted difference in planet size between the models is also true under mixed-envelope scenarios, in which envelopes consist of mixtures of hydrogen and steam. We highlight that transit surveys of young exoplanets should target the youngest observable stellar clusters to exploit the effects of mean molecular weight.
\end{abstract}


%%%%%%%%%%%%%%%%%%%%%%%%%%%%%%%%%%%%%%%%%%%%%%%%%%

%%%%%%%%%%%%%%%%% BODY OF PAPER %%%%%%%%%%%%%%%%%%

\section{Introduction} \label{sec:intro}
Twenty years on from the first sub-Neptune discoveries \citep{Santos2004,Rivera2005}, understanding their nature still remains a mystery \citep[e.g.][]{Bean2021}.  With typical sizes between $\sim2 - 4 R_\oplus$, and orbital periods $\lesssim 100$~days, sub-Neptunes orbit $\sim 50\%$ of Sun-like stars \citep{Petigura2013,Batalha2013,Fressin2013}. Sub-Neptunes sit above the ``radius valley'', an observed paucity of planets at $\sim 1.8R_\oplus$, although its exact location varies with orbital period and stellar mass \citep[e.g.][]{Fulton2017,Petigura2018,VanEylen2018,Petigura2022,Ho2023}. Their counterparts, referred to as super-Earths, sit below the valley. Crucially, mass measurements have shown that the bulk density for planets above and below the valley are markedly different \citep[e.g.][]{Wu_Lithwick2013,Weiss2014,Rogers2015,Wolfgang2016,Chen2017}. Super-Earths are consistent with a rocky, Earth-like composition. Conversely, sub-Neptunes have much lower bulk densities, suggesting the presence of low-density compositional components within these planets. The radius valley can thus also be interpreted as an approximate delineation between planet types of different bulk densities \citep{Luque2022}. Despite the discovery of thousands of sub-Neptunes, their compositions and formation pathways are still highly debated.

Two major hypotheses exist to explain these observations. In the ``gas dwarf'' scenario, super-Earths and sub-Neptunes formed from a single population of progenitor planets with thick primordial H$_2$/He-dominated atmospheres accreted from the nascent protoplanetary disc \citep[e.g.][]{Lee2014,Ginzburg2016}. Then, through atmospheric escape processes, including boil-off \citep{Owen2016,Ginzburg2016}, core-powered mass loss \citep{Ginzburg2018} and XUV photoevaporation \citep{Owen2013,LopezFortney2013}, planets that are smaller in mass and closer to their host stars were stripped of their envelopes, leaving behind bare cores. These rocky cores thus formed the population of super-Earths, whereas the planets that could retain their H$_2$/He envelopes formed the sub-Neptune population. This evolutionary model is supported by direct observations of escaping hydrogen and helium \citep[e.g.][]{DosSantos2023a}.

%Direct evidence for this hypothesis exists with observations of escaping H$_2$/He atmospheres through transit spectroscopy and demographic studies 

On the other hand, the ``water world'' hypothesis states that the radius valley represents the end-product of two distinct planet formation pathways \citep[e.g.][]{Mordasini2009,Venturini2016,Raymond2018,Zeng2019,Mousis2020,Burn2024}. Rocky super-Earths were formed in the inner regions of protoplanetary discs, where the major solid components are silicates. Sub-Neptunes formed beyond the water-ice line, where the available solid mass increases due to the condensation of volatile species. They accrete large quantities of ice/water content (typically $\sim 50\%$ of their mass) and then migrate inwards to the orbital locations we observe them at today.

Seeking evidence in favour of either model at the population level is a complex challenge, largely due to the compositional degeneracy when only measuring a planet's bulk density \citep[e.g.][]{Rogers2015,Bean2021}. Models for both mechanisms can reproduce demographic observations in various parameter spaces, such as orbital period, planet radius, planet mass, and stellar mass spaces \cite[e.g.][]{Gupta2019,Rogers2021,Burn2024}. Water worlds are predicted to be more prominent around M-dwarfs since the water-ice line may be closer to the host star, providing a larger reservoir of water-rich material closer to the location where sub-Neptunes are observed. However, observational evidence for this remains unclear \citep{Rogers2023b,Ho2024}. 

In this paper, we present a new approach to distinguishing between gas dwarfs and water worlds, which relies on the characterisation of planet demographics at early stellar ages. We exploit the difference in predicted mean molecular weights in the envelopes of such planets, which produce very different evolutionary behaviour at young ages. Gas dwarf envelopes are expected to be H$_2$/He-dominated, yielding a relatively low mean-molecular weight and, thus, large scale height. The thickness of this envelope is sensitive to the thermal state of the planet, which will be larger when the planet is young. Thus, gas dwarfs are predicted to be puffy at young ages. Water worlds, on the other hand, are expected to have volatile-rich envelopes with high mean molecular weights and small scale heights. The thickness of such envelopes will respond weakly to changes in thermal state, meaning that they will not undergo significant contraction with time \citep[see also][]{Lopez2017,Aguichine2024}. This provides an observational difference between the models, which can be exploited. Recent atmospheric characterisation of evolved, temperate sub-Neptunes from \textit{JWST} have revealed a wide range in upper atmospheric mean molecular weights \citep[e.g.][]{Madhusudhan2023,Benneke2024,Piaulet-Ghorayeb2024,Davenport2025}. Meanwhile, \textit{JWST}/\textit{HST} observations of young sub-Neptunes have suggested low mean molecular weights \citep[e.g.][]{Barat2024,Thao2024}. However, these observations are sparse, biased towards M-dwarfs, and only probe the upper atmospheres of sub-Neptunes, preventing robust constraints from being placed on formation pathways at the population level.

Our paper is laid out as follows: in Section \ref{sec:modelmethod}, we present a semi-analytical model for gas dwarf and water world structure and evolution, and justify that their radii differ greatly at young ages in Section \ref{sec:MeanMolecularWeight}. In Section \ref{sec:populationlevel}, we compare population-level models with observed planets from \textit{Kepler} and \textit{TESS}. Discussion and conclusions are provided in Sections \ref{sec:discussion} and \ref{sec:conclusion}, respectively.

\section{A simple model for gas dwarfs and water worlds} \label{sec:modelmethod}
The envelopes of close-in, highly irradiated sub-Neptunes are expected to be in radiative-convective equilibrium. To investigate their evolution, we construct semi-analytic models based on previous works for H$_2$/He-dominated envelopes \citep[e.g.][]{Owen2017,Gupta2019,Misener2021} and steam-dominated envelopes \citep{Rogers2025}. Assuming an ideal gas equation of state, the density profile in the convective envelope is assumed to be adiabatic and given by \citep[e.g.][]{Ginzburg2016}:
\begin{equation} \label{eq:temperature_profile_adiabat}
    \rho(r) = \rho_\text{rcb}  \bigg[ 1 + \nabla_\text{ad} \frac{G M_\text{c}}{c_\text{s}^2 R_\text{rcb}} \bigg( \frac{R_\text{rcb}}{r} - 1 \bigg) \bigg]^{\frac{1}{\gamma - 1}},
\end{equation}
where the subscript `rcb' refers to quantities evaluated at the radiative-convective boundary. Here,  $\rho$ is the gas density, $\gamma$ is the adiabatic index, $\nabla_\text{ad} \equiv (\gamma - 1) / \gamma$ is the adiabatic temperature gradient, $M_\text{c}$ is the planet's core mass and $R_\text{rcb}$ is the position of the radiative-convective boundary. Above the convective region sits a radiative layer, assumed to be isothermal at an equilibrium temperature of $T_\text{eq}$. The sound speed in Equation \ref{eq:temperature_profile_adiabat} is evaluated at the radiative-convective boundary, $c_\text{s}^2 \equiv k_\text{B} T_\text{eq} / \mu$, where $\mu$ is the mean molecular weight of the envelope, and is a crucial variable in this study. We highlight here that, for simplicity, we assume the mean molecular weight to be constant in the envelope as a function of radius. We will frequently refer to $\mu$ as the ``bulk'' mean molecular weight, which can be interpreted as a mass-averaged value throughout the envelope. The density profile in this radiative layer is given by:
\begin{equation} \label{eq:temperature_profile_isotherm}
    \rho = \rho_\text{rcb} \exp \bigg\{ \frac{GM_\text{c}}{c_\text{s}^2} \bigg( \frac{1}{r} - \frac{1}{R_\text{rcb}} \bigg) \bigg\}.
\end{equation}
To evolve the model, we numerically solve a set of coupled ordinary differential equations:
\begin{equation} \label{eq:dMdt}
    \frac{d M_\text{env}}{dt} = - \dot{M},
\end{equation}
\begin{equation} \label{eq:dEdt}
    \frac{d E_\text{p}}{dt} = -L_\text{rad}.
\end{equation}
Here, Equation \ref{eq:dMdt} represents the conservation of mass, where $M_\text{env}$ is the envelope mass and $\dot{M}$ is the mass loss rate. Equation \ref{eq:dEdt} represents the conservation of planet energy, $E_\text{p}$.\footnote{Strictly speaking, Equation \ref{eq:dEdt} should have an extra term proportional to the mass loss rate, $\dot{M}$, since escaping material also removes energy from the planet. This is important for extreme mass loss processes such as boil-off \citep[e.g.][]{Owen2016}. However, our evolution models begin after this phase. Hence we can safely ignore this term.} Here, $L_\text{rad}$ is the radiative luminosity of the planet, which, in radiative-convective equilibrium, is given by:
\begin{equation} \label{eq:Lrad}
    L_\text{rad} = \frac{64 \pi \nabla_\text{ad} G M_\text{p}  \sigma T_\text{eq}^4}{3 \, c_\text{s}^2 \, \kappa_\text{rcb} \, \rho_\text{rcb}},
\end{equation}
where $\kappa_\text{rcb}$ is the gas Rosseland mean opacity evaluated at the radiative-convective boundary. 

Calculating the envelope mass $M_\text{env}$ and planetary energy $E_\text{p}$ requires integrating through the planetary structure. For envelope mass:
\begin{equation}
    M_\text{env} = \int_{R_\text{c}}^{R_\text{out}} 4 \pi r^2 \rho(r) dr,
\end{equation}
where $R_\text{c}$ is the core radius and $\rho$ is given by Equations \ref{eq:temperature_profile_adiabat} and \ref{eq:temperature_profile_isotherm}. The outer boundary, $R_\text{out}$ is given by: 
\begin{equation}
    R_\text{out} = \text{min} ( R_\text{B}, R_\text{H} ),
\end{equation}
where $R_\text{B} = G M_\text{c} / 2 c_\text{s}^2$ is the Bondi radius and $R_\text{H} = a (M_\text{c} / 3M_*)^{1/3}$ is the Hill radius. Planetary energy is calculated by:
\begin{equation}
\begin{split}
    E_\text{p} & = \int_{R_\text{c}}^{R_\text{out}} 4 \pi r^2 \rho(r) \, \bigg( \frac{k_\text{B} T(r)}{ \mu \,(\gamma - 1) } - \frac{GM_\text{c}}{r} \bigg) \, dr \\
    & + \frac{M_\text{c} k_\text{B} T_\text{c}}{\mu_\text{c} \, (\gamma_\text{c} - 1)} ,
\end{split}
\end{equation}
where the first and second terms represent energy stored in the envelope and core, respectively. For the core, we follow \citet{Misener2021} and assume an incompressible and isothermal core at a temperature of $T_\text{c}$ with mean molecular weight $\mu_c$ and adiabatic index $\gamma_\text{c} = 4/3$ from the Dulong-Petit law. 

To solve Equations \ref{eq:dMdt} and \ref{eq:dEdt}, we use a \verb|RK45| adaptive-step integrator with a numerical tolerance of $10^{-6}$. At each timestep, the updated envelope mass, $M_\text{env}$, and planet energy, $E_\text{p}$, are used to solve for the new radiative-convective boundary density, $\rho_\text{rcb}$, and radius, $R_\text{rcb}$. We also calculate the position of the photospheric radius, $R_\text{ph}$, and transit radius, $R_\text{tr}$, via \verb|RK45| numerical integration \citep[e.g.][]{Guillot2010}. This is required since the difference between photospheric and transit radii of young planets with large scale heights can become significant $(\gtrsim 10\%) $.

Since we are interested in young planets and their host stars, we include pre-main sequence stellar evolution from \citet{Johnstone2021}. The stellar luminosity, $L_*(t)$, for a given stellar mass and age, is used to calculate planetary equilibrium temperature as follows:
\begin{equation}
    T_\text{eq} = \bigg( \frac{L_*}{16 \sigma \pi a^2} \bigg)^\frac{1}{4},
\end{equation}
where we have assumed a zero Bond albedo. 

For initial conditions, we take a simple, agnostic approach in setting each planet's cooling timescale, $\tau_\text{cool}$, equal to the planet's age, $t_\text{pl}$. The cooling timescale, which is closely related to the Kelvin-Helmholtz timescale used in other works \citep[e.g.][]{Owen2017,Owen2020}, is given by:
\begin{equation} \label{eq:coolingtimescale}
    \tau_\text{cool} = \frac{E_\text{c} - E_\text{env}}{L_\text{rad}},
\end{equation}
where the energy of the envelope $E_\text{env}$ is always negative if the envelope is gravitationally bound. This initial condition of $\tau_\text{cool} = t_\text{pl}$ enforces an initial planet entropy appropriate for its age. In the early stages of planetary evolution, the planet’s thermal state and cooling rate are closely tied to its formation process and contraction. Assuming the cooling timescale is comparable to the planet's age ensures that the initial model reflects a system for which cooling balances with the internal thermal content over the planet’s lifetime. In other words, if $\tau_\text{cool} \ll t_\text{pl}$, the planet will start very inflated and contract unphysically quickly. On the other hand, $\tau_\text{cool} \gg t_\text{pl}$ would imply a very contracted initial size and very little thermal evolution with time, which is likely also unphysical for a recently formed planet. We note here that extreme mass loss events, such as boil-off during protoplanetary disc dispersal, may increase a planet's initial cooling timescale since high entropy material is removed from the planet very quickly, thus introducing premature cooling \citep{Owen2016,Ginzburg2016,Owen2020,Rogers2024a,Tang2024}. To account for this, we also enforce that a planet's initial radiative-convective boundary cannot exceed $0.1 R_\text{B}$, where $R_\text{B} = G M_\text{p} / 2 c_\text{s}^2$ is the Bondi radius. This constraint is appropriate for a planet in which a bolometrically-driven transonic wind, triggered by disc dispersal, has largely shut off \cite[see][]{Owen2016}. These initial conditions are thus appropriate for a planet for which its nascent protoplanetary disc has dispersed and any consequential mass loss has concluded.

In addition to the initial conditions, various model parameters, including the core's mean molecular weight, $\mu_\text{c}$, gas mean molecular weight, $\mu$, gas opacity, $\kappa$, and gas adiabatic index, $\gamma$, depend on the assumed formation model. These are discussed in Sections \ref{sec:GasDwarfParams} and \ref{sec:WaterWorldParams}.

\subsection{Gas Dwarfs} \label{sec:GasDwarfParams}

For our simplest gas dwarf models, we assume a gas mean molecular weight of $\mu = 2.35$~atomic mass units (amu), appropriate for Solar-metallicity gas \citep[e.g.][]{Anders1989}, a gas adiabatic index of $\gamma = 7/5$ for molecular hydrogen, and interpolate the Rosseland mean opacity tables from \citet{Kempton2023} for Solar-metallicity gas. For the cores of gas dwarfs, multiple studies have shown that their interiors are consistent with an Earth-like composition \citep[e.g.][]{Owen2017,Gupta2020,Rogers2021,Rogers2023}. Thus, we assume a core mean molecular weight of $\mu_\text{c} = 60$, motivated by Earth \citep{Szurgot2015}. To calculate the core radius, $R_\text{c}$, we adopt the mass-radius relations of \citet{Fortney2007}, assuming an Earth-like interior composition of $25\%$ iron, $75\%$ rock, as inferred from \citet{Rogers2021,Rogers2023}.

Hydrogen-dominated gas is susceptible to atmospheric escape, which can include mechanisms such as boil-off during disc dispersal \citep{Owen2016,Ginzburg2016,Rogers2024a}, core-powered mass loss \citep[e.g.][]{Ginzburg2018,Gupta2019,Misener2024} and XUV photoevaporation \citep[e.g.][]{OwenJackson2012,Kubyshkina2022,Schulik2022,Caldiroli2022}. For simplicity, we choose to model planets after the protoplanetary disc has fully dissipated. As shown in \citet{Rogers2024a}, boil-off and core-powered mass loss should dominate during disc dispersal, then photoevaporation dictates the planetary outflow once the gas disc is optically thin to XUV photons through its midplane. Nevertheless, photoevaporation and core-powered mass loss are similar hydrodynamic escape processes \citep{Owen2024}, and thus, this simplifying assumption has little effect on our results. We calculate photoevaporative mass loss rates with the following \citep{Lecavelier2007,Erkaev2007}:
\begin{equation} \label{eq:Mdot}
    \dot{M} = \eta \frac{L_\text{XUV}}{4\pi a^2} \frac{\pi R_\text{ph}^3}{GM_\text{p}},
\end{equation}
where $\eta$ is the mass loss efficiency, calculated by interpolating over the grid of hydrodynamic models from \citet{OwenJackson2012}. Typical efficiency values for sub-Neptunes can range between $\sim 10^{-3} - 10^{-1}$, depending on planet mass and radius. The XUV luminosity from the host star, $L_\text{XUV}$, is taken from the pre-main sequence stellar evolution models from \citet{Johnstone2021}.

\subsection{Water Worlds} \label{sec:WaterWorldParams}
For our simplest water world models, we assume a pure steam envelope with a mean molecular weight of $\mu = 18$~amu, a gas adiabatic index of $\gamma = 4/3$, and Rosseland mean opacities from \citet{Kempton2023} for pure H$_2$O gas. For the core, we assume an interior composition of $50\%$ rock, $50\%$ water, as predicted by population synthesis models \citep[e.g.][]{Mordasini2009,Raymond2018,Zeng2019,Venturini2016,Mousis2020,Burn2024}. This yields a core mean molecular weight of $\mu_\text{c}=27$~amu. We use this composition in the mass-radius relations of \citet{Fortney2007} to calculate a core radius, $R_\text{c}$. Similar to the steam models of \citet{Rogers2025}, we choose not to consider the atmospheric escape of steam-dominated (or, more generally, high mean molecular weight) envelopes due to the many uncertainties of this mechanism. Little is currently known about how processes such as molecular dissociation/ionisation, drag, diffusion, and line-cooling may affect the escape of high mean molecular weight envelopes, and thus, we leave this for future work. Nevertheless, studies have shown that water-rich sub-Neptunes should store most of their water in their rocky interiors \citep{Dorn2021,Luo2024}, meaning that we need only consider small envelope mass fractions, which may either suffer from inefficient atmospheric escape and/or may be replenished from their outgassing interiors. 

\section{The impact of envelope mean molecular weight on planet size} \label{sec:MeanMolecularWeight}

\begin{figure}
	\includegraphics[width=\columnwidth]{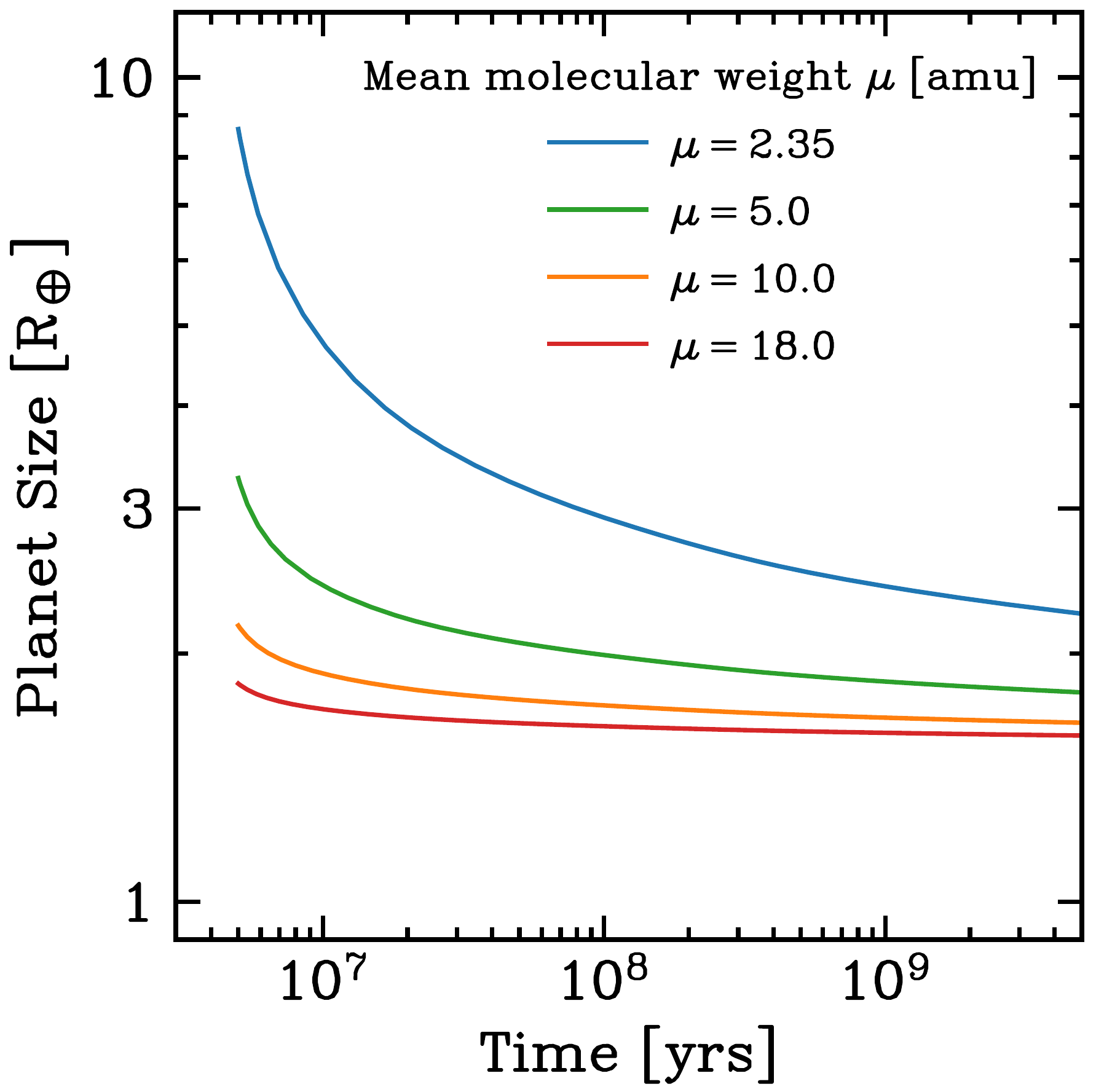}
    \centering
        \cprotect\caption{The transit radius evolution of sub-Neptune envelopes for different bulk mean molecular weights of $\mu = 2.35$, $5.0$, $10$ and $18.0$~amu in blue, green, orange and red, respectively. All planets have a $5M_\oplus$ core, with an Earth-like core composition of $25\%$ iron, $75\%$ silicate. Each model has an envelope mass fraction of $X \equiv M_\text{env} / M_\text{c} = 0.01$ and a constant equilibrium temperature of $T_\text{eq}=800$~K and does not undergo any mass loss.} \label{fig:mmw} 
\end{figure} 

\begin{figure*}
	\includegraphics[width=2.0\columnwidth]{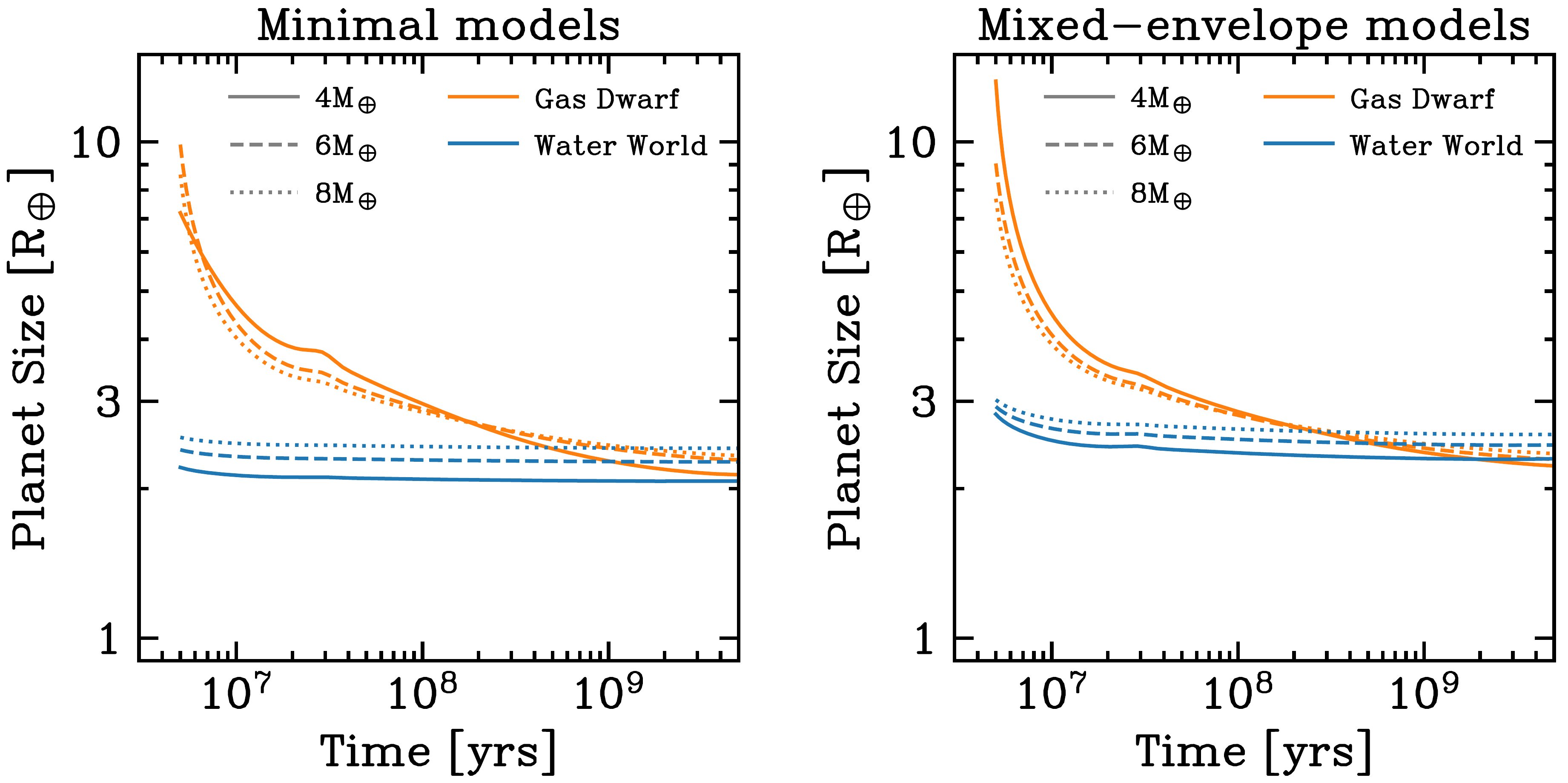}
    \centering
        \cprotect\caption{The radius evolution for different sub-Neptune models: gas dwarfs with Earth-like interiors and H$_2$/He-dominated envelopes (orange) and water worlds with water-rich interiors and steam-dominated atmospheres (blue). Planet size is shown for core masses of $4M_\oplus$, $6M_\oplus$ and $8M_\oplus$ in solid, dashed and dotted, respectively. The left-hand panel shows ``minimal'' models, representing the simplest construction of each planet formation hypothesis. Gas dwarfs have water-poor, Earth-like interiors, host Solar-metallicity envelopes with bulk mean molecular weights of $\mu = 2.35$~amu, initial $5\%$ envelope mass fractions and undergo photoevaporative mass loss. Water worlds have water-rich interiors, host pure steam envelopes with bulk mean molecular weights of $\mu = 18$~amu with envelope mass fractions of $1\%$ and do not undergo mass loss. The right-hand panel shows mixed envelope models, for which gas dwarfs have bulk mean molecular weights of $\mu = 5$~amu and initial envelope mass fractions of $30\%$. Water worlds have bulk mean molecular weights of $\mu = 6$~amu with envelope mass fractions of $1\%$. Despite converging to the same radius after $\sim$~Gyrs of evolution under various assumptions, gas dwarfs and water worlds behave very differently at early times.} \label{fig:Comparison} 
\end{figure*} 

The mean molecular weight of a gaseous envelope directly controls its scale height, $H$. In the simplistic case of an isothermal ideal gas, the scale height is:
\begin{equation}
    H(r) = \frac{c_\text{s}^2}{g} = \frac{ k_\text{B} T}{ \mu} \frac{r^2}{G M_\text{p}} \propto \frac{1}{\mu},
\end{equation}
where $g$ is the gravitational acceleration at radius $r$. The scale height is inversely proportional to the mean molecular weight for a given temperature (or, equivalently, thermal state of the envelope). The mean molecular weight also impacts the scale height for a non-isothermal envelope, such as in the convective interior. From Equation \ref{eq:temperature_profile_adiabat}, one can show that the pressure gradient, which is analogous to the scale height, $|dP / dr| \propto \mu^{1 / (\gamma-1)}$. In other words, the envelope pressure in a convective interior falls off more steeply for a high mean molecular weight composition, implying a less puffy envelope for a given mass and entropy. In summary, two envelopes of the same mass and entropy but with different mean molecular weights will, therefore, be different sizes. 

As a sketch of the idea, Figure \ref{fig:mmw} demonstrates the evolution of a sub-Neptune for different bulk mean molecular weights, ranging from $\mu=2.35$~amu (appropriate for Solar metallicity H$_2$/He-dominated gas) to $\mu=18.0$~amu (appropriate for a pure H$_2$O gas). All planets have a $5M_\oplus$ core, with an interior Earth-like core composition. Each model has an envelope mass fraction of $X \equiv M_\text{env} / M_\text{c} = 0.01$. For simplicity, we assume a constant equilibrium temperature of $T_\text{eq}=800$~K (ignoring the pre-main sequence evolution of the host star for now), a constant gas opacity of $\kappa = 0.1 \text{ cm}^2 \text{ g}^{-1}$ and do not include any mass loss. Whereas the size of planets converge after Gyrs of evolution, they differ significantly at early times. The high temperatures from formation increase the size of planets with lower mean molecular weight more substantially, causing them to be more inflated when young. In this work, we highlight that a population of sufficiently young transiting exoplanets would allow one to probe atmospheric bulk mean molecular weights, which are dictated by planet formation and evolution.

\subsection{Gas dwarf vs. water world evolution}

We now consider the evolution of gas dwarfs and water worlds. Figure \ref{fig:Comparison} demonstrates their evolution in orange and blue, respectively. We show planet models with core masses of $4M_\oplus$, $6M_\oplus$ and $8M_\oplus$ in solid, dashed and dotted lines, respectively. Models are initialised at $t=5$~Myrs. In the left-hand panel, we construct ``minimal'' models, as described in Sections \ref{sec:GasDwarfParams} and \ref{sec:WaterWorldParams}, which represent the simplest physical construction of each formation scenario. Under this set of models, gas dwarfs and water worlds have bulk mean molecular weights of $\mu=2.35$~amu and $\mu=18.0$~amu, respectively, initial envelope mass fractions of $X = 0.05$ and $X = 0.01$, respectively, and orbital periods of $10$~days. Gas dwarfs and water worlds converge in size after $\sim 1$~Gyr due to their internal composition producing the same radius: water worlds with water-rich, low-density cores and compact envelopes, gas dwarfs with rocky/iron-rich, high-density cores and extended envelopes. However, their evolution is divergent at early ages. The low bulk mean molecular weight of gas dwarf envelopes means that the energy of formation, which is abundant at early times, puffs the planet up significantly when compared to water worlds. This is particularly apparent for planet ages $\lesssim 100$~Myrs.

In the right-hand panel of Figure \ref{fig:Comparison}, we consider more complex models of gas dwarfs and water worlds, informed by recent theoretical and observational studies. In the case of gas dwarfs, multiple works have shown that the chemical interaction of Solar-metallicity gas with a magma ocean at high temperatures will result in an increase in envelope mean molecular weight \citep{Kite2021,Schlichting2022,Misener2022,Misener2023,Rogers2024b,Young2024}. Most notably, chemical reactions between hydrogen gas and molten silicate can produce large quantities of vapour species such as SiO, MgO, FeO and H$_2$O, increasing the mean molecular weight above Solar values. As an example, recent \textit{JWST} observations of TOI-270 d, an archetypal temperate sub-Neptune, have been interpreted with a mean molecular weight of $\mu \sim 5$~amu \citep{Benneke2024}, consistent with the global chemical equilibrium models of \citet{Schlichting2022,Rogers2024b}. To capture this effect in our simple evolutionary formalism, we produce gas dwarf models with bulk mean molecular weights of $\mu=5.0$~amu, which is indicative of mixed miscible envelopes. Since H$_2$O is a dominant opacity carrier, we use the Rosseland mean opacities from \citet{Kempton2023} for a H$_2$/He/H$_2$O mixture with a H$_2$O mass fraction of $60\%$ to reproduce this bulk mean molecular weight. We continue to assume atmospheric escape according to Equation \ref{eq:Mdot} with the same efficiencies as in the minimal gas dwarf model, but highlight that molecular cooling and drag should reduce such mass loss rates. This model thus provides an upper limit in terms of contraction rate as a test of whether the mixed-envelope models can still be distinguished at early ages. In order for this model to reproduce the typical sizes of sub-Neptunes, we must increase its envelope mass fraction, as highlighted in \citet{Thorngren2019,Misener2022}. For the right-hand panel of Figure \ref{fig:Comparison}, gas dwarfs have an initial envelope mass fraction of $X=0.3$. Note that it is beyond the scope of this work to model the convection-inhibition brought about from mean molecular weight gradients in the lower envelope, as also highlighted in \citet{Misener2022,Misener2023}. Our goal is to demonstrate the population-level impacts of bulk mean molecular weight on the sizes of young sub-Neptunes. Nevertheless, convection inhibition can alter the thermal evolution of planets, and should be considered in future work. 

Similar to our mixed-envelope models of gas dwarfs, the right-hand panel of Figure \ref{fig:Comparison} also shows water worlds with H$_2$/He/H$_2$O mixed envelopes. Recent population synthesis models from \citet{Burn2024, Burn2024b} have considered the accretion of solar-metallicity gas onto water-rich sub-Neptunes. They predict envelope water mass fractions as low as $\sim 70\%$, equating to a mean molecular weight of $\mu \approx 6$~amu, which we adopt in our mixed-envelope models of water worlds. We use the Rosseland mean opacities from \citet{Kempton2023} for a H$_2$/He/H$_2$O mixture with a H$_2$O mass fraction of $70\%$. We retain an atmospheric mass fraction of $X = 0.01$ from our ``minimal'' models, motivated by theoretical studies of water-rich planets hosting most of their H$_2$O in the interior \citep{Luo2024}. Recent \textit{JWST} upper-atmospheric characterisation of GJ 9827 d reveals an abundance of H$_2$O and a mean molecular weight between $\mu \sim 10$ and $18$~amu \citep{Piaulet-Ghorayeb2024}. This planet is thus consistent with some predicted water world scenarios.

% While transit/emission spectroscopy from \textit{JWST} may shed light on the mean molecular weights in the upper regions of sub-Neptune atmospheres, they do not probe the bulk atmospheric composition. Furthermore, these observations are sparse due to the expense and limited number of suitable targets. 

Regardless of the underlying assumptions, Figure \ref{fig:Comparison} highlights that the radius evolution of gas dwarfs and water worlds are dramatically different at early ages. One can also see that planet size is relatively insensitive to planet mass as a function of time. This can be understood from a simple virial theorem argument, which states that, for a spherically-symmetric envelope in hydrostatic equilibrium, its gravitational potential energy, $U = - GM_\text{c} M_\text{env} / R_\text{p}$, is related to its internal energy, $E_\text{i}$:
\begin{equation}
    U = -3(\gamma-1) E_\text{i},
\end{equation}
where we have safely ignored vanishing boundary terms at large radii and assumed an adiabatic equation of state for gas pressure, $P\propto \rho^\gamma$. The internal energy for an ideal gas envelope scales as $E_\text{i} \propto \bar{P}_\text{env}V_\text{env}$, where $V_\text{env}$ is the volume of the atmosphere and $\bar{P}_\text{env}$ is a mass-averaged envelope pressure \citep[e.g.][]{kippenhahn2012stellar}. In the case of gas dwarfs with non-self-gravitating envelopes, particularly at young ages, the volume is dominated by the radius of the envelope, hence $V_\text{env} \propto R_\text{p}^3$ and the pressure can be approximated as $\bar{P}_\text{env} \propto (M_\text{env} / R_\text{p}^3)^\gamma$ from the adiabatic equation of state. Combining, and recalling that $M_\text{env} \equiv XM_\text{c}$, one finds that:
\begin{equation} \label{eq:Rp_Vs_X_Mc}
    R_\text{p} \propto X^\frac{2}{3} \, M_\text{c}^{-\frac{1}{3}}.
\end{equation}
In other words, the radius of a planet is insensitive to its mass for a given atmospheric mass fraction and age.\footnote{Equation \ref{eq:Rp_Vs_X_Mc} is derived assuming an adiabatic index of $\gamma = 5/3$, since $\gamma=4/3$ provides an analytic singularity. When the full semi-analytic structure equations are numerically solved, as in Figure \ref{fig:Comparison}, one recovers an insensitive planet size as a function of planet mass for a fixed envelope mass fraction.} This is well-documented in other works \citep[see Figure 1 from][]{Lopez2014}.

\section{Young transiting exoplanets as a probe of planet formation at the population level} \label{sec:populationlevel}

In Section \ref{sec:MeanMolecularWeight} we have demonstrated that the radius evolution of sub-Neptune envelopes can vary substantially with different bulk mean molecular weights. We now build population-level models under the two categories, gas dwarfs vs. water worlds, to showcase how such scenarios would present at early times. For simplicity, we only construct population models from the ``minimal'' formation scenarios since these are better studied when compared to mixed-envelope models as described in Sections \ref{sec:GasDwarfParams} and \ref{sec:WaterWorldParams}. As such, we assume that gas dwarfs and water worlds have bulk envelope mean molecular weights of $\mu = 2.35$~amu and $18.0$~amu, respectively, for the remainder of this paper. Nevertheless, Figure \ref{fig:Comparison} demonstrates that more complex mixed-envelope models also show similar differences in radius evolution at early times. 

\subsection{Modelling synthetic transit surveys} \label{sec:survey}
The prerequisite for our population models is that they reproduce the observed exoplanet population at late times $\gtrsim 1$~Gyr. To do so, we choose to compare the models at late times with the \textit{California Kepler Survey} (CKS) from \citet{Petigura2022}. We seek to construct underlying models that, when injected through the \textit{CKS} pipeline, reproduce the observed (biased) population. For each model, we evolve $10^6$ planets, with stellar masses drawn from a Gaussian distribution with a mean of $1M_\odot$ and a standard deviation of $0.3M_\odot$ \citep{Rogers2021b}. We also randomly draw a disc dispersal time, which defines the initial time for each planet model, uniformly drawn between $3$~Myrs and $10$~Myrs \citep[e.g.][]{Kenyon1995,Ercolano2011,Koepferl2013}. The typical ages of host stars in the \textit{CKS} catalogue are in the range of $\approx 1-10$~Gyrs \citep[e.g.][]{Berger2020}. Hence, we terminate each planet model at a random time, uniformly drawn between these limits. Finally, we statistically model observational uncertainty in planetary radii by introducing a random Gaussian perturbation to each planet radius with zero mean and a standard deviation of $5\%$ the planet's true size.

We produce synthetic \textit{CKS} surveys at late times by randomly drawing $25,200$ planets from the underlying population of $10^6$ modelled planets. This value comes from an assumed occurrence rate of $0.7$ planets per star in the \textit{CKS} stellar catalogue of $36,000$ stars \citep{Fulton2017}. Thus, the expectation value for planets orbiting the surveyed stars is $0.7 \times 36,000 = 25,200$. For each planet, we calculate the probability of transit and detection following \citet{Fulton2017,Rogers2021,Petigura2022}. The probability of transit is given by:
\begin{equation}
    p_\text{tr} = \frac{R_*}{a},
\end{equation}
where $R_*$ is the stellar radius and $a$ is the semi-major axis. The probability of detection, $p_\text{det}$, takes the form of a $\Gamma$ cumulative distribution function of the form:
\begin{equation} \label{eq:pdet}
    p_\text{det} = A \, \Gamma(k) \int^{\frac{m-l}{\theta}}_{0} t^{k-1} \; e^{-t} \; \mathrm{d} t,
\end{equation}
where $A=1.0$, $k=17.56$, $l=1.00$ and $\theta=-0.49$ following \citet{Fulton2017}. The signal-to-noise ratio, $m$, for a transiting planet is:
\begin{equation} \label{eq:SNR}
   m = \bigg( \frac{R_\text{p}}{R_*}\bigg)^2 \; \sqrt \frac{T_\text{obs}}{P} \; \frac{1}{\text{CDPP}},
\end{equation}
where $T_\text{obs}=4$~years is the observation time for \textit{Kepler}, $P$ is the orbital period, and the combined differential photometric precision (CDPP) is a measure of noise for the host star, measured in ppm. To match the results from \citet{Koch2010}, we randomly draw $\log (\text{CDPP})$ as a Gaussian distribution with a mean and standard deviation of $-4.0$ and $0.25$, respectively. The product of $p_\text{tr} \times p_\text{det}$ is the probability that a given planet will be observed in a transit survey. For each planet, we draw a random number between $0$ and $1$. If this number is less than $p_\text{tr} \times p_\text{det}$, then this planet is ``observed'' in the survey. For each sub-Neptune formation model, we repeat this process to produce $1000$ synthetic surveys. We then use a kernel density estimator (KDE) to calculate a probability density function (PDF) for planet detection in period-radius space from the combined surveys. This PDF is then scaled to the integrated expectation value for survey yield, i.e. the number of detected planets in each modelled survey. An underlying formation model is a good fit to the data if this expectation value approximately matches the actual yield of $926$ planets in the \textit{CKS} catalogue, as well as the distribution shape of observed planets in period-radius space.

Given a population-level model that reproduces the late-time demographics, the next step is to inspect the model predictions at early times. We choose to compare these models with a small sample of observed transiting exoplanets from \textit{TESS} \citep{Vach2024a}. Figure \ref{fig:Comparison} highlights that the gas dwarf and water world models diverge the most at the earliest times, hence we only include planets within this sample with host stellar ages $<40$~Myrs.\footnote{We also update this sample to include the recently detected $3$~Myr old planet from \citet{Barber2024}.} We evaluate each modelled planet radius at a random time, uniformly drawn up to $40$~Myrs and again add a radius measurement uncertainty of $5\%$. In the \textit{TESS} survey of \citet{Vach2024a}, the stellar sample for stars $<40$~Myrs contains $\approx 2500$ stars. In order to produce synthetic surveys, we again assume the same occurrence rate of $0.7$ planets per star, and draw $1750$ planets from our underlying population of young planets. Since \textit{Kepler} and \textit{TESS} have different completeness and noise properties, we approximate the probability of detection for \textit{TESS} with the same functional form as in Equation \ref{eq:pdet} with $A=0.85$, $k=5.0$, $l=1.00$ and $\theta=2.5$, which best reproduces Figure 8 from \citet{Vach2024a} for young FGK stars. Similarly, the signal-to-noise ratio for \textit{TESS} transits is calculated with Equation \ref{eq:SNR} with $T_\text{obs}=60$~days. Although \textit{TESS} does not quantify photometric precision with a CDPP, we find an effective $\log (\text{CDPP})$ randomly drawn from a Gaussian distribution with a mean and standard deviation of $-1.5$ and $0.25$, respectively, best reproduces completeness maps from \citet{Fernandes2022,Vach2024a}. As with the \textit{CKS} synthetic surveys, we repeat this entire procedure $1000$ times and produce a combined PDF for planet detection, scaled to the integrated expectation value of synthetic survey yield. This is compared to the $14$ planets with hosts stars $< 40$~Myrs in the \citet{Vach2024a} survey.

We stress that comparing models at early evolutionary times is only valid if the population models reproduce the observations at late times. In order to reproduce the \textit{CKS} sample at $\sim 1 - 10$~Gyrs, both population-level models require different planetary initial conditions, including initial envelope mass fraction distributions, core mass distributions, and orbital period distributions. We discuss these in the following sections.

\subsection{Gas dwarf population model}

% We stress that this model predicts that both super-Earths and sub-Neptunes formed through the same formation pathway and then separated due to atmospheric escape. Hence, all planets in this model are given the same core composition.

For the gas dwarf case, both super-Earths and sub-Neptunes are drawn from the same underlying population, and atmospheric escape is responsible for separating the two sub-populations over time and forming the radius valley. We use the inferred underlying distributions from \citet{Rogers2021}, who fitted the XUV photoevaporation ``minimal'' model from Section \ref{sec:GasDwarfParams} to the \textit{CKS} data of \citet{Fulton2018} using a Bayesian hierarchical model. Both the core mass and initial envelope mass fraction distributions are parameterised using Bernstein polynomials of $5^{\text{th}}$ order. These are peaked distributions, with centres at $\sim 4M_\oplus$ and $\sim 2\%$, respectively. The orbital period distribution is parameterised with a smooth broken power-law as follows:
\begin{equation} \label{eq:Porb_dist}
    \frac{dN}{dP} \propto \frac{1}{(\frac{P}{P_0})^{-k_1} + (\frac{P}{P_0})^{-k_2}},
\end{equation}
where $P_0 = 5.5$~days, $k_1 = 2.3$, and $k_2=0.0$. We assume all cores have a composition of $25\%$ iron, $75\%$ rock \citep{Rogers2021,Rogers2023} with Solar-metallicity H$_2$/He envelopes with mean-molecular weights of $\mu=2.35$~amu. Again, we stress that models that include self-consistent interior-envelope chemical coupling are still in their infancy \citep[e.g.][]{Kite2021,Schlichting2022,Misener2023}. They are also computationally expensive, meaning that population modelling is currently challenging. While we leave these considerations for future work, Figure \ref{fig:Comparison} highlights that the difference in early evolution between gas dwarfs and water worlds under this scenario is still significant, meaning our results will remain largely the same. 

\subsection{Water world population model}

The fundamental prediction of the water world hypothesis is that super-Earths and sub-Neptunes formed through different channels. Therefore, we require different underlying distributions for the two classes of planets. For simplicity, we choose not to model these planets during the disc phase, as done in population-synthesis modelling \citep[e.g.][]{Burn2024}, which includes large-scale disc migration to bring water worlds interior to the water-ice line. Instead, we produce appropriate underlying distributions for the post-disc evolution phase that reproduce the \textit{CKS} data at $\sim$~Gyr times. To match the ratio of super-Earths and sub-Neptunes in these observations, $35\%$ of the underlying planet population are modelled as super-Earths, $65\%$ as sub-Neptunes.

For the core mass distribution, we adopt a log-Gaussian distribution:
\begin{equation}
    \frac{dN}{d\,\log M_\text{c}} \propto \exp \bigg\{ -\frac{(\log M_\text{c} - \log \mu_\text{M})^2}{2\sigma_\text{M}^2}\bigg\}.
\end{equation}
For super-Earths, we set the mean and standard deviation to $\mu_\text{M} = 3M_\oplus$ and $\sigma_\text{M} = 0.35$, respectively. For sub-Neptunes, we set $\mu_\text{M} = 10M_\oplus$ and $\sigma_\text{M} = 0.30$, respectively.

For the orbital period distributions, we use Equation \ref{eq:Porb_dist} with $P_0 = 4.5$~days, $k_1 = 2.0$, and $k_2=-0.5$ for super-Earths, and $P_0 = 8.5$~days, $k_1 = 2.5$, and $k_2=0.1$ for sub-Neptunes. These values are all consistent with the \textit{CKS} study of \citet{Petigura2022}.

As discussed, the water world hypothesis predicts that super-Earths and sub-Neptunes have different core compositions due to their different formation channels. We assume Earth-like compositions of $25\%$ iron, $75\%$ rock for super-Earths with no atmospheres \citep{Rogers2021,Rogers2025}. For sub-Neptunes, we adopt a core composition of $50\%$ water, $50\%$ rock and draw a random steam envelope mass fraction log-uniformly between $0.001\%$ and $10\%$ \citep{Luo2024}.

\subsection{Population-level results}

\begin{figure*}
	\includegraphics[width=2.0\columnwidth]{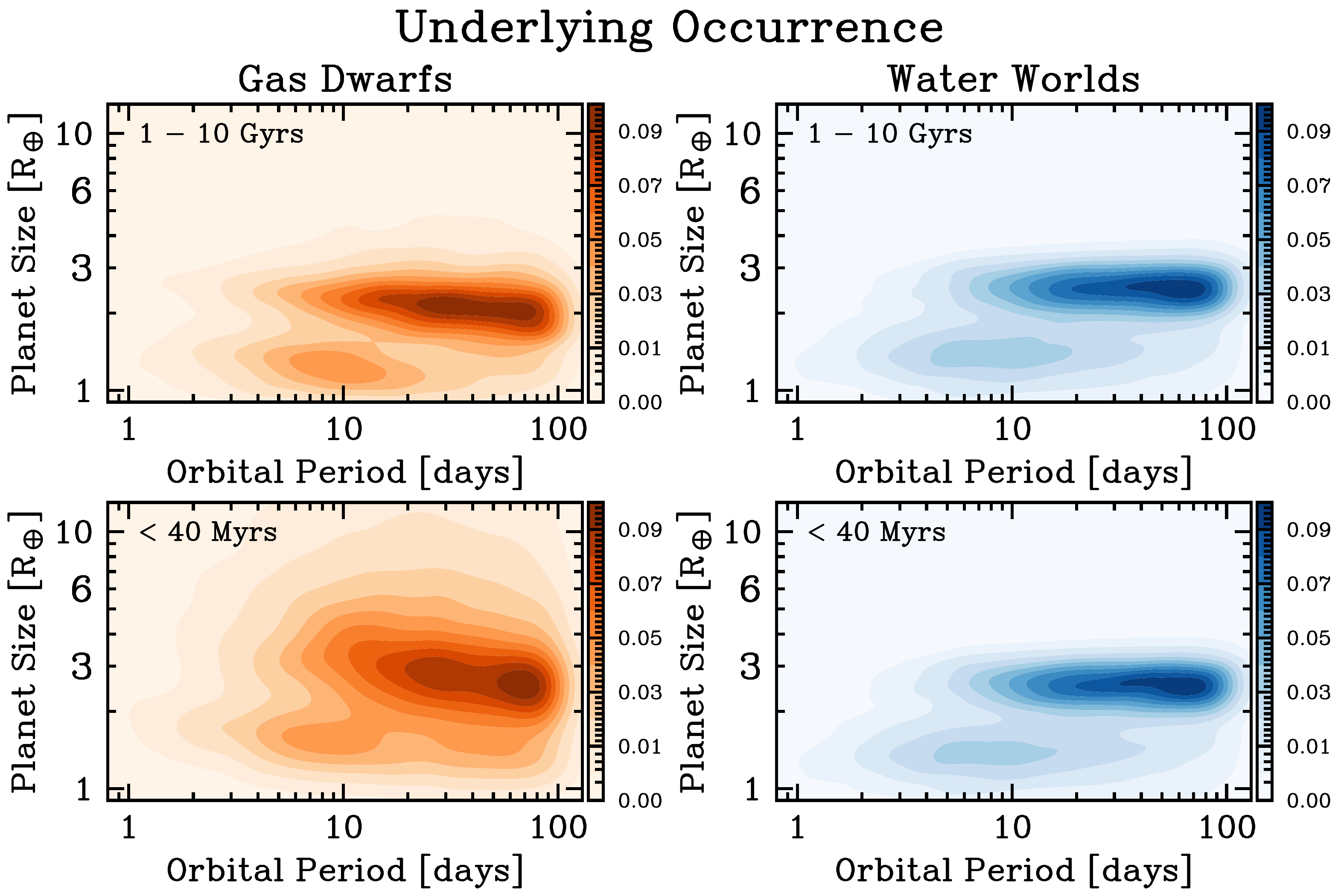}
    \centering
        \cprotect\caption{Population-level predictions of gas dwarf and water world hypotheses for sub-Neptune evolution in the left and right-hand panels, respectively. The populations are shown at times of $1-10$~Gyrs and $<40$~Myrs in the top and bottom panels, respectively. Colour bars represent relative probability of planet occurrence.} \label{fig:PopulationComparison_Occ} 
\end{figure*} 

\begin{figure*}
	\includegraphics[width=2.0\columnwidth]{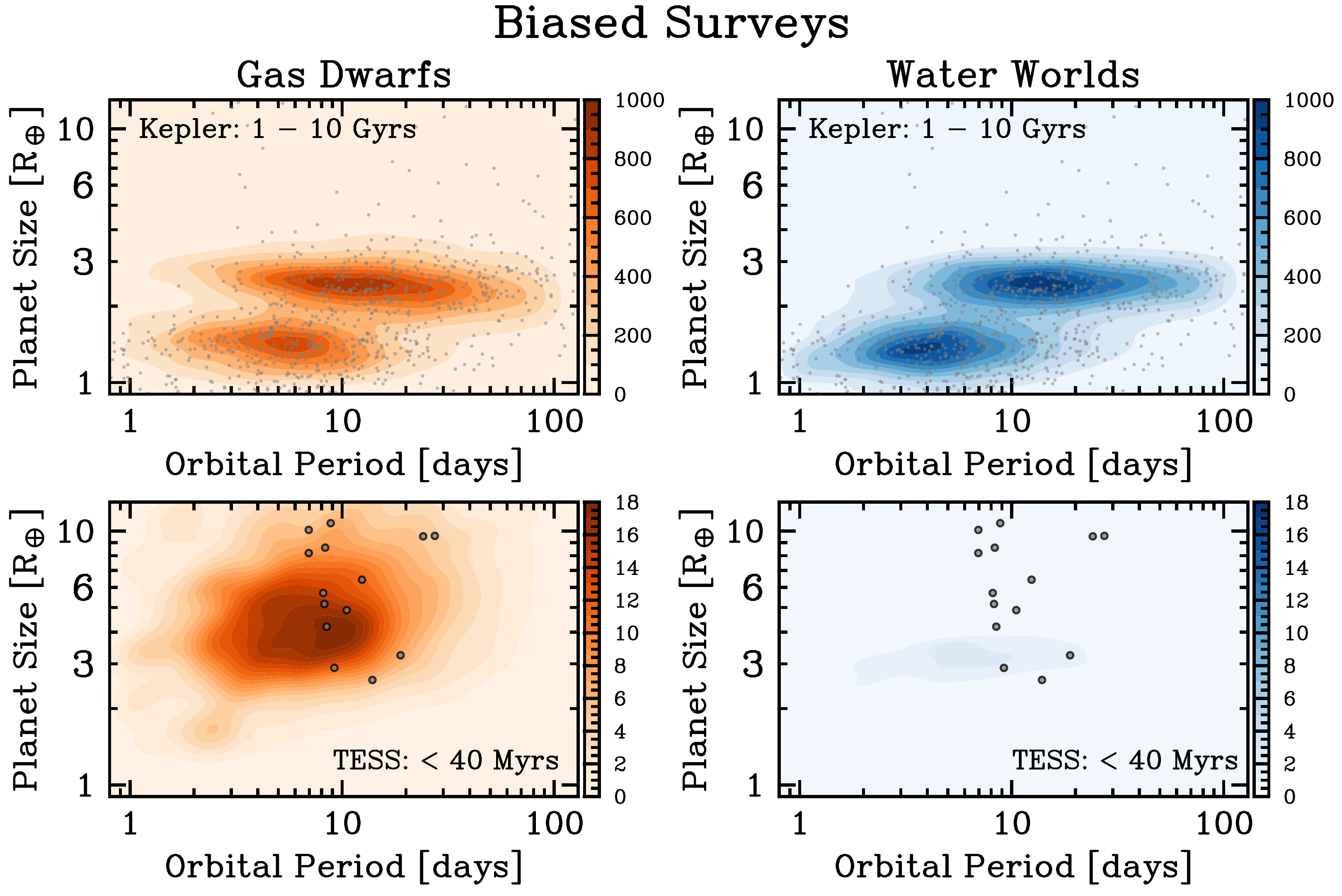}
    \centering
        \cprotect\caption{Same as Figure \ref{fig:PopulationComparison_Occ} but now accounting for survey biases and uncertainty. Colour bars represent the integrated expectation value for survey yield. We perform synthetic \textit{California Kepler Surveys} on each model in the top panels, with the real observed population shown as grey points \citep{Petigura2022}. Similarly, we perform synthetic \textit{TESS} surveys for host stars $<40$~Myrs in the bottom panels, with real observed planets shown as grey points \citep{Vach2024a}.} \label{fig:PopulationComparison_Det} 
\end{figure*} 

The underlying unbiased planet populations for gas dwarf and water worlds are presented in the left and right-hand panels of Figure \ref{fig:PopulationComparison_Occ}, respectively. These are presented at late times ($1-10$~Gyrs) and early times ($<40$~Myrs) in the top and bottom panels, respectively. Colour bars represent the relative probability of planet occurrence. One can see the emergence of the radius valley in the gas dwarf case, with the ratio of super-Earths to sub-Neptunes increasing with time as atmospheric escape strips primordial sub-Neptunes into bare rocky super-Earths. The typical sizes of gas dwarf sub-Neptunes also significantly decrease with time as a result of their low bulk mean molecular weights. The water world scenario, on the other hand, shows very little evolution since the super-Earths and sub-Neptunes formed separately in the disc phase, and the water world sub-Neptunes have high bulk mean molecular weight, meaning very little thermal contraction.

Figure \ref{fig:PopulationComparison_Det}, on the other hand, shows the predicted biased survey yields at late and early times, as described in Section \ref{sec:survey}. These are compared to the \textit{CKS} exoplanet sample from \citet{Petigura2022}, and the young \textit{TESS} sample from \citet{Vach2024a}. In this case, colour bars represent the expectation value for integrated survey yield. Both models reproduce the observations at late times, with two distinct sub-populations of super-Earths and sub-Neptunes, separated by the radius valley. The expectation value for survey yield is $910^{+20}_{-16}$ for gas dwarfs, and $931^{+24}_{-14}$ for water worlds, which are both consistent with the real \textit{CKS} survey yield of $926$.

The nature of \textit{TESS}' lower photometric precision and the higher activity of young stars means that the expected yields of such surveys are considerably lower than \textit{Kepler}. The survey completeness is significantly lower for planets smaller than $\sim 3 R_\oplus$, meaning that super-Earths are very unlikely to be detected under either model. As an example, a planet of radius $3 R_\oplus$ at $0.1$~AU orbiting a main-sequence $1 M_\odot$ star is predicted to have a signal-to-noise ratio of $m \sim 100$ from Equation \ref{eq:SNR} as detected by \textit{Kepler}, but $m \sim 5$ if orbiting a young active star detected by \textit{TESS}. This equates to a probability of detection from Equation \ref{eq:pdet} of $p_\text{det} \sim 1$ and $p_\text{det} \sim 0.07$, respectively. As a result of this systematic reduction in detectability for young active stars observed by \textit{TESS}, the expectation value for survey yield under the gas dwarf model is significantly reduced to $15^{+3}_{-2}$ planets. The actual yield from \textit{TESS} is $14$ planets $< 40$~Myrs in \citet{Vach2024a}. For the water world model, the the expected yield is $3^{+3}_{-1}$ planets, which is inconsistent with \textit{TESS} at the $\sim 2-3\sigma$ level. The gas dwarf model is better at reproducing the observations since these planets are larger at younger ages due to their lower bulk mean molecular weights, meaning they are more likely to be detected. One can see that the gas dwarf model predicts planet detections at significantly larger radii at early times, as seen in the \textit{TESS} data.

Although the young transiting exoplanet sample from \textit{TESS} is visually more consistent with the gas dwarf model, it is interesting to note that the predicted orbital period distribution is noticeably different to the data. Young \textit{TESS} planets pile up at $\gtrsim 10$~days, despite completeness being significantly higher at shorter orbital periods. This specific realisation of the gas dwarf model thus over-predicts the number of sub-Neptunes at short orbital periods at young ages. The majority of these predicted close-in sub-Neptunes would eventually lose their atmospheres to become super-Earths on $\sim 100$~Myr to $\sim$~Gyr timescales. The discrepancy could thus be caused by an under-prediction of planets that were ``born rocky'', i.e. began life as super-Earths in the post-disc epoch. Alternatively, it could be due to post-disc inward orbital migration that changes the period distribution with time. We discuss this in Section \ref{sec:super-Earths}. Although not as statistically significant, it is also worth noting that the gas dwarf model slightly under-predicts the number of large sub-Neptunes $\sim 10$R$_\oplus$ at young ages. This could be explained by higher initial entropies of formation or perhaps delayed radial contraction due to convection inhibition in the envelope of such planets \citep[e.g.][]{Misener2022}, which we discuss in Section \ref{sec:theory_uncertainties}.

A full statistical comparison between models and data is beyond the scope of this work, and requires Bayesian model comparison that takes into account the many free parameters of each model. In other words, there are likely realisations of both models that reproduce the data marginally better than those presented in this work, hence a comparison is not fair unless these realisations are fully explored. Model comparison involves calculating a Bayes factor via the posterior distributions of both models, which is a computationally expensive exercise and hence left for future work when the young \textit{TESS} sample has increased in size. We highlight that the sample of young transiting exoplanets is small, and thus current inferences are likely dominated by small number statistics. Nevertheless, various studies have shown a tentative overabundance of large sub-Neptunes ($\gtrsim 4 R_\oplus$) orbiting young stars, which may hint at low mean-molecular weight, gas dwarf evolution \citep[e.g.][]{Fernandes2022,Christiansen2023,Vach2024a,Fernandes2025}. 

\section{Discussion} \label{sec:discussion}
We have shown that the bulk mean molecular weight of a small exoplanet envelope directly affects its thermal contraction with time. At early times this can result in dramatically different evolution depending on the assumed formation model. In particular, we have focussed on the gas dwarf and water world hypotheses regarding the formation of sub-Neptunes. We now discuss caveats to this strategy, and an observational and theoretical route to determine the nature of sub-Neptunes at the population level.

\subsection{Gas opacities and cooling timescales} \label{sec:Opacity}

\begin{figure}
	\includegraphics[width=\columnwidth]{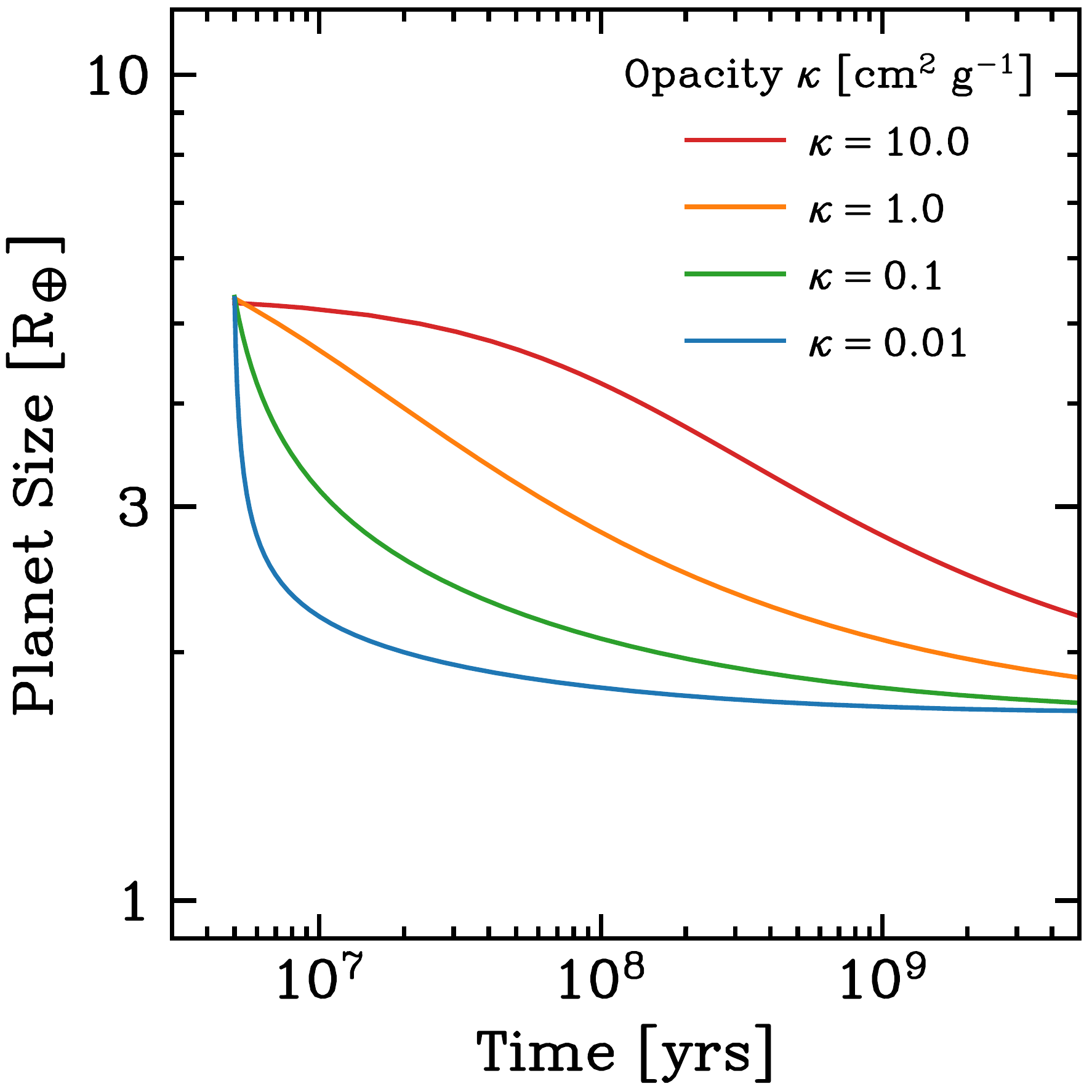}
    \centering
        \cprotect\caption{The radius evolution of exoplanet atmospheres for different Rosseland mean gas opacities of $\kappa = 0.01$, $0.1$, $1.0$ and $10.0$ cm$^2$ g$^{-1}$ in blue, green, orange and red, respectively. All planets have a $5M_\oplus$ core, with an Earth-like core composition of $25\%$ iron, $75\%$ silicate. Each model has an atmospheric mass fraction of $X \equiv M_\text{atm} / M_\text{c} = 0.01$ and a constant equilibrium temperature of $T_\text{eq}=800$~K. All planets are initialised with a radiative-convective boundary set at $5 R_\oplus$.} \label{fig:kappa} 
\end{figure} 

An important envelope property that controls the evolution of a planet is gas opacity. Whereas the mean molecular weight influences the instantaneous size of an envelope (via the scale height), the gas opacity also affects the instantaneous \textit{rate of change} of envelope size. From Equation \ref{eq:Lrad}, one can see that the radiative luminosity, which controls the cooling and contraction of a planet, is inversely proportional to the opacity. Thus, higher gas opacities yield slower thermal contraction. As a demonstration of this, Figure \ref{fig:kappa} shows the evolution of the same planet from Figure \ref{fig:mmw}, however, here, we vary the gas Rosseland mean opacity from $0.01 \text{ cm}^2 \text{ g}^{-1}$ to $10.0 \text{ cm}^2 \text{ g}^{-1}$ with a mean molecular weight of $\mu = 5.0$~amu. To highlight the difference in evolution, we set the initial radiative-convective boundary radii to $R_\text{rcb} = 5 R_\oplus$. Again, planets converge to the same size after several Gyrs, but their initial cooling rate is altered by varying opacities. One can also see that the planets with the lowest opacities contract with a timescale much shorter than their age (seen by near-vertical lines in Figure \ref{fig:kappa}). This suggests that the initial condition of $R_\text{rcb} = 5 R_\oplus$ was likely inappropriate for these envelopes. The probability of detecting such a planet at its initially large size is thus vanishingly small, given its rapid contraction. Since we have observed several young sub-Neptunes at young ages with large radii $\sim 10 \,R_\oplus$, it follows from a statistical argument that their cooling timescales must be at least equal to their ages. In our work, we chose to adopt agnostic initial conditions in which the cooling timescale of the planet is equal to its age, which in turn sets the initial size of the planet at the point of disc dispersal. More work is needed to investigate the initial conditions of planets under different scenarios, as discussed in Section \ref{sec:theory_uncertainties}.

To avoid the potential degeneracy between gas opacity and mean molecular weight when comparing models with observations of young transiting planets, one should use self-consistent models that couple the two properties. For example, a steam-dominated envelope will have larger mean opacities than solar metallicity H$_2$/He-dominated gas. This coupling is also true for the adiabatic index, $\gamma$, which in this work parametrises the equation of state within a convective envelope (see Equation \ref{eq:temperature_profile_adiabat}) and can also vary depending on the species present in an atmosphere. The adiabatic index is important since it controls the size of a planet's convective envelope for a given entropy. For example, a lower adiabatic index, implying more degrees of freedom for a species, reduces the size of a convective envelope for a fixed entropy. This effect, however, is smaller in magnitude than that of changing the bulk mean molecular weight in the envelope. For our simplistic models, we have used the opacities for the appropriate mean molecular weight and H$_2$/He/H$_2$O mixture, thus maintaining self-consistency within the model. Nevertheless, future studies should include full opacity and equation of state tables, which account for intricate changes in chemical properties as the planet evolves. Improvement to model sophistication in this manner will help in refining the difference in the size of gas dwarfs and water worlds at young ages. 

If one wishes to exploit the effects of bulk mean molecular weight, an important observational strategy is to observe planets at the youngest possible age. This minimises the number of cooling timescales a planet has evolved over, and thus captures the state of each planet closer to its initial conditions. The major reason that we are unable to infer the underlying nature of sub-Neptunes at $\sim$~Gyr timescales is that they have evolved for many cooling timescales, destroying the information of their individual formation pathway. Recent detections of very young transiting exoplanets $<10$~Myrs \citep[e.g.][]{Barber2024} have shown that these detections are indeed possible and offer exciting glimpses into newly-formed planets.

\subsection{On the benefits of probing the young super-Earth population} \label{sec:super-Earths}

In Figure \ref{fig:PopulationComparison_Det}, we see that the gas dwarf formation model is visually better at reproducing the young planet population than the water world model. However, one can also see that the gas dwarf model over-predicts the number of detected young sub-Neptunes at short orbital periods $\lesssim 10$~days when compared to the \textit{TESS} data. One plausible explanation for this discrepancy is an under-prediction of planets that are ``born rocky'' within the gas dwarf model. \citet{Rogers2021} used \textit{CKS} data to infer that $\lesssim 20\%$ of planets should be born as super-Earths by the end of disc dispersal, which may be explained with a combination of processes, including inefficient gas accretion, boil-off and core-powered mass loss \citep[e.g.][]{Lee2022,Owen2024,Rogers2024a}. Explicitly modelling these processes will likely reduce the number of predicted sub-Neptunes at short orbital periods at early times, and thus allow the model to better reproduce the observations. We leave this increase in model complexity for future work. An alternative explanation for the lack of detected \textit{TESS} planets $\lesssim 10$~days is that the orbital period distribution of sub-Neptunes evolves during the post-disc era. Inward migration of sub-Neptunes and super-Earths could be caused by the interaction of their escaping atmospheres with stellar winds, resulting in a net angular momentum loss for the planet \citep{Hanf2025}, or due to tidal migration \citep[e.g.][]{LeeOwen2025}.

One way to determine which of these mechanisms is responsible for the lack of observed young sub-Neptunes at short orbital periods is to also measure the orbital period distribution of super-Earths. If there is an abundance of super-Earths at short orbital periods at early times, it is likely that inefficient gas accretion and/or atmospheric escape produced this population during the disc phase. Characterising the young super-Earth population can also be used to measure the ratio of super-Earths to sub-Neptunes as a function of time. One clear prediction of the gas dwarf model is that this ratio should increase with time since sub-Neptunes are being stripped of their atmospheres to become super-Earths \citep[e.g.][]{Gupta2020,Rogers2021}. However, probing this area of parameter space $(R_\text{p} \lesssim 2 R_\oplus)$ around young stars is technically challenging for instruments such as \textit{TESS}, and will require next-generation transit observations.

\subsection{An observational route forward}
More detections of young transiting exoplanets are required to answer science questions regarding planet formation and evolution at the population level. Previous studies have shown that the young planet population is inconsistent with the evolved \textit{Kepler} population at the $\sim 1-3\sigma$ level \citep{Fernandes2022,Christiansen2023,Vach2024a,Fernandes2025}, with a strong indication that young sub-Neptunes are more inflated, and more consistent with gas dwarf models. These studies have been limited, however, by various observational effects, including the heightened activity of pre-main sequence host stars and the requirement of high-resolution imaging to resolve stars within their dense clusters. Recall from Section \ref{sec:Opacity} that the most informative planet detections are those from the youngest stellar clusters, which are limited in number and magnitude for missions such as \textit{TESS}. 

Mission concepts, such as the \textit{Early eVolution Explorer (EVE)}, aim to perform dedicated transit surveys of young stellar clusters \citep{Howard2024}. With the increase in planet yield from this style of survey, one could also investigate mixture models in which planets from both populations are present. This statistical analysis lends itself to a Bayesian framework, allowing one to infer the appropriate fraction of each model that best fits the data. 

As shown in Section \ref{sec:MeanMolecularWeight}, the radius of an inflated planet is insensitive to planet mass, which conveniently means that well-characterised planet masses are not necessary to make inferences from population models. Recent work has also shown that most young transiting exoplanets are very unlikely to be progenitor hot-Jupiters \citep{Karalis2024}. Nevertheless, transit spectroscopy of young planet atmospheres from \textit{JWST}/\textit{HST} can constrain a planet's mass from its scale height \citep[e.g.][]{Barat2024,Thao2024}. Atmospheric characterisation may, in fact, be a more straightforward route to a young planet's mass than conventional methods, such as radial velocity measurements, which are notoriously challenging to obtain around young active stars. This also comes with the obvious benefit of constraints on atmospheric chemistry at early times. \textit{EVE} thus also serves as a path-finder for suitable young planet targets for atmospheric characterisation with \textit{JWST} and future missions such as \textit{ARIEL} and \textit{HWO}.

\subsection{Theoretical uncertainties} \label{sec:theory_uncertainties}
In this work, we have utilised simple radiative-convective equilibrium models to demonstrate that the bulk mean molecular weight of an envelope alters its size most noticeably at young ages. However, these models have neglected some important physical effects, which should be closely examined in future work. 

The models in this study have assumed a constant mean molecular weight as a function of envelope radius. As discussed in Section \ref{sec:MeanMolecularWeight}, recent theoretical work has shown that planet envelopes chemically react with their interiors, which in turn changes the mean molecular weight within the atmosphere \citep[e.g.][]{Kite2021,Schlichting2022}. Mean molecular weight gradients can form, which may inhibit convection in the deep interior and thus change the envelope structure and evolution \citep[e.g.][]{Leconte2017,Misener2022}. The effects of phase equilibria and miscibility between various species such as H$_2$, H$_2$O, and silicate melt also remain largely unexplored \citep[e.g.][]{Mousis2020,Vazan2022,Pierrehumbert2023,Gupta2024,Young2024}. Coupled models of planet interiors and envelopes must be considered, particularly for young planets due to their higher entropies and thus active geochemistry. As discussed in Section \ref{sec:populationlevel}, the gas dwarf population model presented in this work slightly under-predicts the number of large sub-Neptunes $\sim 10$~R$_\oplus$ at young ages. More sophisticated evolution modelling of such effects may help in interpreting this under-prediction.   

Beyond the interior and envelope structure of a planet, more attention should be given to understanding the initial conditions of planet evolution. For gas dwarf models (as discussed in Section \ref{sec:GasDwarfParams}), studies have shown that the rapid disc dispersal process in the inner regions $\lesssim 1$~AU induces an extreme mass loss event referred to as ``boil-off'', in which a planet may lose $\sim 90$~\% of its atmospheric mass \citep[e.g.][]{Owen2016,Ginzburg2016,Rogers2024a,Tang2024}. These studies have shown that the initial cooling timescale of a gas dwarf post-disc dispersal may be longer than its age, which is caused by extreme cooling of the envelope during boil-off \citep{Owen2020}. In this study, we accounted for boil-off by enforcing initial planet sizes not exceeding $0.1 R_\text{B}$ \citep[see][]{Owen2016}. Nevertheless, in future studies, the initial conditions of sub-Neptunes under various formation and composition models should be investigated. In addition, more work is needed to understand the mass loss of high mean molecular weight species in exoplanet atmospheres. This is a complex challenge due to various species-dependant heating and cooling processes, as well as drag and diffusion.

Finally, the key to statistically inferring properties from a population of young transiting exoplanets is to use self-consistent structure and evolution models that can reproduce the properties of evolved exoplanets at $\sim$~Gyr epochs. These must also be computationally fast in order to run large quantities of population-level models. The balance between model complexity and speed may be aided with machine-learnt emulators \citep{Rogers2023}.

\section{Conclusions} \label{sec:conclusion}
In this paper, we have discussed the effects of bulk envelope mean molecular weight on the early evolution of close-in exoplanets. The lower the bulk mean molecular weight, the larger the radius of a planet at early times, which provides an observational window into their composition and, thus, formation pathway. We have focussed on sub-Neptunes, which can currently be explained with the ``gas dwarf'' hypothesis (rocky interiors and extended H$_2$/He dominated envelopes) or the ``water world'' hypothesis (rocky interiors with several tens of per cent in water mass fraction). A population of young transiting exoplanets will allow for model comparison to determine which of these models best describes reality. Our conclusions are as follows:

\begin{itemize}
    \item The observed size of a young sub-Neptune is sensitive to the bulk mean molecular weight of its envelope. The lower the mean molecular weight, the larger the scale height, and thus the larger the radius inflation. However, as planets cool, they contract. As a result, the range in planet radii as a function of mean molecular weight is greatest at earliest times (Figure \ref{fig:mmw}).
    \vspace{0.2cm}
    \item The radii of evolved sub-Neptunes can be explained by two interior compositions: "gas dwarfs" with low global water mass fractions and low envelope mean molecular weights in the approximate range of $\mu \sim 2.35-5$~amu, or "water worlds" with high water mass fractions and high envelope mean molecular weights in the range of $\mu \sim 6 - 18$~amu. The radii of sub-Neptunes under these formation pathways converge after $\sim$~Gyrs of evolution. However, their radii diverge significantly at early ages $\lesssim 100$~Myrs (Figure \ref{fig:Comparison}).
    \vspace{0.2cm}
    \item The prediction of different planet radii for the gas dwarf and water world hypotheses can be exploited with a population of young transiting exoplanets in order to probe mean molecular weight and, hence, sub-Neptune composition. We show that surveys of young planets should target the youngest observable stellar clusters to maximise this difference to distinguish between formation channels.
    \vspace{0.2cm}
    \item For young, puffy sub-Neptunes, planet size is relatively insensitive to planet mass. This removes the necessity of characterising planet masses to infer bulk composition, which is notoriously difficult for planets orbiting young, active stars. Nevertheless, atmospheric spectroscopy may provide a clearer path to determining the masses of young planets as shown in recent works \citep[e.g.][]{Thao2024}.
    \vspace{0.2cm}
    \item There is tentative evidence that the population of observed transiting planets from \textit{TESS} \citep{Vach2024a} is inconsistent with a simple water world model (Figure \ref{fig:PopulationComparison_Det}). However, this sample is small and, thus, the results may be driven by small-number statistics. We encourage future statistical comparisons to place emphasis on accounting for this effect, as well as other survey biases in forward models, for a fair comparison.   
\end{itemize}

Inferring the composition of sub-Neptunes at the population level will inevitably rely on theoretical models. Future studies should place emphasis on understanding the implications of chemical interactions between the interior and envelopes of such planets, as well as possible initial conditions in their post-disc phase under a range of formation scenarios.

\section*{Acknowledgements}
I would like to thank the anonymous referee for valuable comments that helped improve this paper. I would also like to thank Sydney Vach for sharing the \textit{TESS} planet sample used in this study, as well as Jennifer Burt, Neal Turner, Andrew Mann, Madyson Barber, George Zhou, Eric Gaidos, Hilke Schlichting and James Owen for discussions that helped improve this paper. JGR gratefully acknowledges support from the Kavli Foundation.
%%%%%%%%%%%%%%%%%%%%%%%%%%%%%%%%%%%%%%%%%%%%%%%%%%
\section*{Data Availability}

All models will be made available upon reasonable request.



\bibliographystyle{mnras}
\bibliography{references} % if your bibtex file is called example.bib

\begin{thebibliography}{}
\makeatletter
\relax
\def\mn@urlcharsother{\let\do\@makeother \do\$\do\&\do\#\do\^\do\_\do\%\do\~}
\def\mn@doi{\begingroup\mn@urlcharsother \@ifnextchar [ {\mn@doi@} {\mn@doi@[]}}
\def\mn@doi@[#1]#2{\def\@tempa{#1}\ifx\@tempa\@empty \href {http://dx.doi.org/#2} {doi:#2}\else \href {http://dx.doi.org/#2} {#1}\fi \endgroup}
\def\mn@eprint#1#2{\mn@eprint@#1:#2::\@nil}
\def\mn@eprint@arXiv#1{\href {http://arxiv.org/abs/#1} {{\tt arXiv:#1}}}
\def\mn@eprint@dblp#1{\href {http://dblp.uni-trier.de/rec/bibtex/#1.xml} {dblp:#1}}
\def\mn@eprint@#1:#2:#3:#4\@nil{\def\@tempa {#1}\def\@tempb {#2}\def\@tempc {#3}\ifx \@tempc \@empty \let \@tempc \@tempb \let \@tempb \@tempa \fi \ifx \@tempb \@empty \def\@tempb {arXiv}\fi \@ifundefined {mn@eprint@\@tempb}{\@tempb:\@tempc}{\expandafter \expandafter \csname mn@eprint@\@tempb\endcsname \expandafter{\@tempc}}}

\bibitem[\protect\citeauthoryear{{Aguichine}, {Batalha}, {Fortney}, {Nettelmann}, {Owen}  \& {Kempton}}{{Aguichine} et~al.}{2024}]{Aguichine2024}
{Aguichine} A.,  {Batalha} N.,  {Fortney} J.~J.,  {Nettelmann} N.,  {Owen} J.~E.,   {Kempton} E. M.~R.,  2024, \mn@doi [arXiv e-prints] {10.48550/arXiv.2412.17945}, \href {https://ui.adsabs.harvard.edu/abs/2024arXiv241217945A} {p. arXiv:2412.17945}

\bibitem[\protect\citeauthoryear{{Anders} \& {Grevesse}}{{Anders} \& {Grevesse}}{1989}]{Anders1989}
{Anders} E.,  {Grevesse} N.,  1989, \mn@doi [\gca] {10.1016/0016-7037(89)90286-X}, \href {https://ui.adsabs.harvard.edu/abs/1989GeCoA..53..197A} {53, 197}

\bibitem[\protect\citeauthoryear{{Barat} et~al.,}{{Barat} et~al.}{2024}]{Barat2024}
{Barat} S.,  et~al., 2024, \mn@doi [Nature Astronomy] {10.1038/s41550-024-02257-0}, \href {https://ui.adsabs.harvard.edu/abs/2024NatAs...8..899B} {8, 899}

\bibitem[\protect\citeauthoryear{{Barber} et~al.,}{{Barber} et~al.}{2024}]{Barber2024}
{Barber} M.~G.,  et~al., 2024, \mn@doi [\nat] {10.1038/nature17448}, \href {https://ui.adsabs.harvard.edu/abs/2024arXiv241118683B} {533, 221}

\bibitem[\protect\citeauthoryear{{Batalha} et~al.,}{{Batalha} et~al.}{2013}]{Batalha2013}
{Batalha} N.~M.,  et~al., 2013, \mn@doi [\apjs] {10.1088/0067-0049/204/2/24}, \href {https://ui.adsabs.harvard.edu/abs/2013ApJS..204...24B} {204, 24}

\bibitem[\protect\citeauthoryear{{Bean}, {Raymond}  \& {Owen}}{{Bean} et~al.}{2021}]{Bean2021}
{Bean} J.~L.,  {Raymond} S.~N.,   {Owen} J.~E.,  2021, \mn@doi [Journal of Geophysical Research (Planets)] {10.1029/2020JE006639}, \href {https://ui.adsabs.harvard.edu/abs/2021JGRE..12606639B} {126, e06639}

\bibitem[\protect\citeauthoryear{{Benneke} et~al.,}{{Benneke} et~al.}{2024}]{Benneke2024}
{Benneke} B.,  et~al., 2024, \mn@doi [arXiv e-prints] {10.48550/arXiv.2403.03325}, \href {https://ui.adsabs.harvard.edu/abs/2024arXiv240303325B} {p. arXiv:2403.03325}

\bibitem[\protect\citeauthoryear{{Berger}, {Huber}, {Gaidos}, {van Saders}  \& {Weiss}}{{Berger} et~al.}{2020}]{Berger2020}
{Berger} T.~A.,  {Huber} D.,  {Gaidos} E.,  {van Saders} J.~L.,   {Weiss} L.~M.,  2020, arXiv e-prints, \href {https://ui.adsabs.harvard.edu/abs/2020arXiv200514671B} {p. arXiv:2005.14671}

\bibitem[\protect\citeauthoryear{{Burn}, {Mordasini}, {Mishra}, {Haldemann}, {Venturini}, {Emsenhuber}  \& {Henning}}{{Burn} et~al.}{2024a}]{Burn2024}
{Burn} R.,  {Mordasini} C.,  {Mishra} L.,  {Haldemann} J.,  {Venturini} J.,  {Emsenhuber} A.,   {Henning} T.,  2024a, \mn@doi [arXiv e-prints] {10.48550/arXiv.2401.04380}, \href {https://ui.adsabs.harvard.edu/abs/2024arXiv240104380B} {p. arXiv:2401.04380}

\bibitem[\protect\citeauthoryear{{Burn}, {Bali}, {Dorn}, {Luque}  \& {Grimm}}{{Burn} et~al.}{2024b}]{Burn2024b}
{Burn} R.,  {Bali} K.,  {Dorn} C.,  {Luque} R.,   {Grimm} S.~L.,  2024b, arXiv e-prints, \href {https://ui.adsabs.harvard.edu/abs/2024arXiv241116879B} {p. arXiv:2411.16879}

\bibitem[\protect\citeauthoryear{{Caldiroli}, {Haardt}, {Gallo}, {Spinelli}, {Malsky}  \& {Rauscher}}{{Caldiroli} et~al.}{2022}]{Caldiroli2022}
{Caldiroli} A.,  {Haardt} F.,  {Gallo} E.,  {Spinelli} R.,  {Malsky} I.,   {Rauscher} E.,  2022, \mn@doi [\aap] {10.1051/0004-6361/202142763}, \href {https://ui.adsabs.harvard.edu/abs/2022A&A...663A.122C} {663, A122}

\bibitem[\protect\citeauthoryear{{Chen} \& {Kipping}}{{Chen} \& {Kipping}}{2017}]{Chen2017}
{Chen} J.,  {Kipping} D.,  2017, \mn@doi [\apj] {10.3847/1538-4357/834/1/17}, \href {https://ui.adsabs.harvard.edu/abs/2017ApJ...834...17C} {834, 17}

\bibitem[\protect\citeauthoryear{{Christiansen} et~al.,}{{Christiansen} et~al.}{2023}]{Christiansen2023}
{Christiansen} J.~L.,  et~al., 2023, \mn@doi [\aj] {10.3847/1538-3881/acf9f9}, \href {https://ui.adsabs.harvard.edu/abs/2023AJ....166..248C} {166, 248}

\bibitem[\protect\citeauthoryear{{Davenport} et~al.,}{{Davenport} et~al.}{2025}]{Davenport2025}
{Davenport} B.,  et~al., 2025, arXiv e-prints, \href {https://ui.adsabs.harvard.edu/abs/2025arXiv250101498D} {p. arXiv:2501.01498}

\bibitem[\protect\citeauthoryear{{Dorn} \& {Lichtenberg}}{{Dorn} \& {Lichtenberg}}{2021}]{Dorn2021}
{Dorn} C.,  {Lichtenberg} T.,  2021, \mn@doi [\apjl] {10.3847/2041-8213/ac33af}, \href {https://ui.adsabs.harvard.edu/abs/2021ApJ...922L...4D} {922, L4}

\bibitem[\protect\citeauthoryear{{Dos Santos}}{{Dos Santos}}{2023}]{DosSantos2023a}
{Dos Santos} L.~A.,  2023, \mn@doi [IAU Symposium] {10.1017/S1743921322004239}, \href {https://ui.adsabs.harvard.edu/abs/2023IAUS..370...56D} {370, 56}

\bibitem[\protect\citeauthoryear{{Ercolano}, {Clarke}  \& {Hall}}{{Ercolano} et~al.}{2011}]{Ercolano2011}
{Ercolano} B.,  {Clarke} C.~J.,   {Hall} A.~C.,  2011, \mn@doi [\mnras] {10.1111/j.1365-2966.2010.17473.x}, \href {http://adsabs.harvard.edu/abs/2011MNRAS.410..671E} {410, 671}

\bibitem[\protect\citeauthoryear{{Erkaev}, {Kulikov}, {Lammer}, {Selsis}, {Langmayr}, {Jaritz}  \& {Biernat}}{{Erkaev} et~al.}{2007}]{Erkaev2007}
{Erkaev} N.~V.,  {Kulikov} Y.~N.,  {Lammer} H.,  {Selsis} F.,  {Langmayr} D.,  {Jaritz} G.~F.,   {Biernat} H.~K.,  2007, \mn@doi [\aap] {10.1051/0004-6361:20066929}, \href {https://ui.adsabs.harvard.edu/abs/2007A&A...472..329E} {472, 329}

\bibitem[\protect\citeauthoryear{{Fernandes} et~al.,}{{Fernandes} et~al.}{2022}]{Fernandes2022}
{Fernandes} R.~B.,  et~al., 2022, arXiv e-prints, \href {https://ui.adsabs.harvard.edu/abs/2022arXiv220603989F} {p. arXiv:2206.03989}

\bibitem[\protect\citeauthoryear{{Fernandes} et~al.,}{{Fernandes} et~al.}{2025}]{Fernandes2025}
{Fernandes} R.~B.,  et~al., 2025, \mn@doi [arXiv e-prints] {10.48550/arXiv.2503.10856}, \href {https://ui.adsabs.harvard.edu/abs/2025arXiv250310856F} {p. arXiv:2503.10856}

\bibitem[\protect\citeauthoryear{{Fortney}, {Marley}  \& {Barnes}}{{Fortney} et~al.}{2007}]{Fortney2007}
{Fortney} J.~J.,  {Marley} M.~S.,   {Barnes} J.~W.,  2007, \mn@doi [\apj] {10.1086/512120}, \href {https://ui.adsabs.harvard.edu/abs/2007ApJ...659.1661F} {659, 1661}

\bibitem[\protect\citeauthoryear{{Fressin} et~al.,}{{Fressin} et~al.}{2013}]{Fressin2013}
{Fressin} F.,  et~al., 2013, \mn@doi [\apj] {10.1088/0004-637X/766/2/81}, \href {https://ui.adsabs.harvard.edu/abs/2013ApJ...766...81F} {766, 81}

\bibitem[\protect\citeauthoryear{{Fulton} \& {Petigura}}{{Fulton} \& {Petigura}}{2018}]{Fulton2018}
{Fulton} B.~J.,  {Petigura} E.~A.,  2018, \mn@doi [\aj] {10.3847/1538-3881/aae828}, \href {https://ui.adsabs.harvard.edu/abs/2018AJ....156..264F} {156, 264}

\bibitem[\protect\citeauthoryear{{Fulton} et~al.,}{{Fulton} et~al.}{2017}]{Fulton2017}
{Fulton} B.~J.,  et~al., 2017, \mn@doi [\aj] {10.3847/1538-3881/aa80eb}, \href {https://ui.adsabs.harvard.edu/abs/2017AJ....154..109F} {154, 109}

\bibitem[\protect\citeauthoryear{{Ginzburg}, {Schlichting}  \& {Sari}}{{Ginzburg} et~al.}{2016}]{Ginzburg2016}
{Ginzburg} S.,  {Schlichting} H.~E.,   {Sari} R.,  2016, \mn@doi [\apj] {10.3847/0004-637X/825/1/29}, \href {http://adsabs.harvard.edu/abs/2016ApJ...825...29G} {825, 29}

\bibitem[\protect\citeauthoryear{{Ginzburg}, {Schlichting}  \& {Sari}}{{Ginzburg} et~al.}{2018}]{Ginzburg2018}
{Ginzburg} S.,  {Schlichting} H.~E.,   {Sari} R.,  2018, \mn@doi [\mnras] {10.1093/mnras/sty290}, \href {https://ui.adsabs.harvard.edu/abs/2018MNRAS.476..759G} {476, 759}

\bibitem[\protect\citeauthoryear{{Guillot}}{{Guillot}}{2010}]{Guillot2010}
{Guillot} T.,  2010, \mn@doi [\aap] {10.1051/0004-6361/200913396}, \href {https://ui.adsabs.harvard.edu/abs/2010A&A...520A..27G} {520, A27}

\bibitem[\protect\citeauthoryear{{Gupta} \& {Schlichting}}{{Gupta} \& {Schlichting}}{2019}]{Gupta2019}
{Gupta} A.,  {Schlichting} H.~E.,  2019, \mn@doi [\mnras] {10.1093/mnras/stz1230}, \href {https://ui.adsabs.harvard.edu/abs/2019MNRAS.487...24G} {487, 24}

\bibitem[\protect\citeauthoryear{{Gupta} \& {Schlichting}}{{Gupta} \& {Schlichting}}{2020}]{Gupta2020}
{Gupta} A.,  {Schlichting} H.~E.,  2020, \mn@doi [\mnras] {10.1093/mnras/staa315}, \href {https://ui.adsabs.harvard.edu/abs/2020MNRAS.493..792G} {493, 792}

\bibitem[\protect\citeauthoryear{{Gupta}, {Stixrude}  \& {Schlichting}}{{Gupta} et~al.}{2024}]{Gupta2024}
{Gupta} A.,  {Stixrude} L.,   {Schlichting} H.~E.,  2024, \mn@doi [arXiv e-prints] {10.48550/arXiv.2407.04685}, \href {https://ui.adsabs.harvard.edu/abs/2024arXiv240704685G} {p. arXiv:2407.04685}

\bibitem[\protect\citeauthoryear{{Hanf}, {Kincaid}, {Schlichting}, {Cappiello}  \& {Tamayo}}{{Hanf} et~al.}{2025}]{Hanf2025}
{Hanf} B.,  {Kincaid} W.,  {Schlichting} H.,  {Cappiello} L.,   {Tamayo} D.,  2025, \mn@doi [\aj] {10.3847/1538-3881/ad944f}, \href {https://ui.adsabs.harvard.edu/abs/2025AJ....169...19H} {169, 19}

\bibitem[\protect\citeauthoryear{{Ho} \& {Van Eylen}}{{Ho} \& {Van Eylen}}{2023}]{Ho2023}
{Ho} C. S.~K.,  {Van Eylen} V.,  2023, \mn@doi [\mnras] {10.1093/mnras/stac3802}, \href {https://ui.adsabs.harvard.edu/abs/2023MNRAS.519.4056H} {519, 4056}

\bibitem[\protect\citeauthoryear{{Ho}, {Rogers}, {Van Eylen}, {Owen}  \& {Schlichting}}{{Ho} et~al.}{2024}]{Ho2024}
{Ho} C. S.~K.,  {Rogers} J.~G.,  {Van Eylen} V.,  {Owen} J.~E.,   {Schlichting} H.~E.,  2024, \mn@doi [arXiv e-prints] {10.48550/arXiv.2401.12378}, \href {https://ui.adsabs.harvard.edu/abs/2024arXiv240112378H} {p. arXiv:2401.12378}

\bibitem[\protect\citeauthoryear{{Howard} et~al.,}{{Howard} et~al.}{2024}]{Howard2024}
{Howard} W.~S.,  et~al., 2024, \mn@doi [arXiv e-prints] {10.48550/arXiv.2411.08092}, \href {https://ui.adsabs.harvard.edu/abs/2024arXiv241108092H} {p. arXiv:2411.08092}

\bibitem[\protect\citeauthoryear{{Johnstone}, {Bartel}  \& {G{\"u}del}}{{Johnstone} et~al.}{2021}]{Johnstone2021}
{Johnstone} C.~P.,  {Bartel} M.,   {G{\"u}del} M.,  2021, \mn@doi [\aap] {10.1051/0004-6361/202038407}, \href {https://ui.adsabs.harvard.edu/abs/2021A&A...649A..96J} {649, A96}

\bibitem[\protect\citeauthoryear{{Karalis}, {Lee}  \& {Thorngren}}{{Karalis} et~al.}{2024}]{Karalis2024}
{Karalis} A.,  {Lee} E.~J.,   {Thorngren} D.~P.,  2024, \mn@doi [arXiv e-prints] {10.48550/arXiv.2408.16793}, \href {https://ui.adsabs.harvard.edu/abs/2024arXiv240816793K} {p. arXiv:2408.16793}

\bibitem[\protect\citeauthoryear{{Kempton}, {Lessard}, {Malik}, {Rogers}, {Futrowsky}, {Ih}, {Marounina}  \& {Mu{\~n}oz-Romero}}{{Kempton} et~al.}{2023}]{Kempton2023}
{Kempton} E. M.~R.,  {Lessard} M.,  {Malik} M.,  {Rogers} L.~A.,  {Futrowsky} K.~E.,  {Ih} J.,  {Marounina} N.,   {Mu{\~n}oz-Romero} C.~E.,  2023, \mn@doi [\apj] {10.3847/1538-4357/ace10d}, \href {https://ui.adsabs.harvard.edu/abs/2023ApJ...953...57K} {953, 57}

\bibitem[\protect\citeauthoryear{{Kenyon} \& {Hartmann}}{{Kenyon} \& {Hartmann}}{1995}]{Kenyon1995}
{Kenyon} S.~J.,  {Hartmann} L.,  1995, \mn@doi [\apjs] {10.1086/192235}, \href {https://ui.adsabs.harvard.edu/abs/1995ApJS..101..117K} {101, 117}

\bibitem[\protect\citeauthoryear{Kippenhahn, Weigert  \& Weiss}{Kippenhahn et~al.}{2012}]{kippenhahn2012stellar}
Kippenhahn R.,  Weigert A.,   Weiss A.,  2012, Stellar Structure and Evolution.
Astronomy and Astrophysics Library, Springer Berlin Heidelberg, \url {https://books.google.co.uk/books?id=wdSFB4B_pMUC}

\bibitem[\protect\citeauthoryear{{Kite} \& {Schaefer}}{{Kite} \& {Schaefer}}{2021}]{Kite2021}
{Kite} E.~S.,  {Schaefer} L.,  2021, \mn@doi [\apjl] {10.3847/2041-8213/abe7dc}, \href {https://ui.adsabs.harvard.edu/abs/2021ApJ...909L..22K} {909, L22}

\bibitem[\protect\citeauthoryear{{Koch} et~al.,}{{Koch} et~al.}{2010}]{Koch2010}
{Koch} D.~G.,  et~al., 2010, \mn@doi [\apjl] {10.1088/2041-8205/713/2/L79}, \href {https://ui.adsabs.harvard.edu/abs/2010ApJ...713L..79K} {713, L79}

\bibitem[\protect\citeauthoryear{{Koepferl}, {Ercolano}, {Dale}, {Teixeira}, {Ratzka}  \& {Spezzi}}{{Koepferl} et~al.}{2013}]{Koepferl2013}
{Koepferl} C.~M.,  {Ercolano} B.,  {Dale} J.,  {Teixeira} P.~S.,  {Ratzka} T.,   {Spezzi} L.,  2013, \mn@doi [\mnras] {10.1093/mnras/sts276}, \href {http://adsabs.harvard.edu/abs/2013MNRAS.428.3327K} {428, 3327}

\bibitem[\protect\citeauthoryear{{Kubyshkina} \& {Fossati}}{{Kubyshkina} \& {Fossati}}{2022}]{Kubyshkina2022}
{Kubyshkina} D.,  {Fossati} L.,  2022, arXiv e-prints, \href {https://ui.adsabs.harvard.edu/abs/2022arXiv221110166K} {p. arXiv:2211.10166}

\bibitem[\protect\citeauthoryear{{Lecavelier Des Etangs}}{{Lecavelier Des Etangs}}{2007}]{Lecavelier2007}
{Lecavelier Des Etangs} A.,  2007, \mn@doi [\aap] {10.1051/0004-6361:20065014}, \href {https://ui.adsabs.harvard.edu/abs/2007A&A...461.1185L} {461, 1185}

\bibitem[\protect\citeauthoryear{{Leconte}, {Selsis}, {Hersant}  \& {Guillot}}{{Leconte} et~al.}{2017}]{Leconte2017}
{Leconte} J.,  {Selsis} F.,  {Hersant} F.,   {Guillot} T.,  2017, \mn@doi [\aap] {10.1051/0004-6361/201629140}, \href {https://ui.adsabs.harvard.edu/abs/2017A&A...598A..98L} {598, A98}

\bibitem[\protect\citeauthoryear{{Lee} \& {Owen}}{{Lee} \& {Owen}}{2025}]{LeeOwen2025}
{Lee} E.~J.,  {Owen} J.~E.,  2025, \mn@doi [\apjl] {10.3847/2041-8213/adafa3}, \href {https://ui.adsabs.harvard.edu/abs/2025ApJ...980L..40L} {980, L40}

\bibitem[\protect\citeauthoryear{{Lee}, {Chiang}  \& {Ormel}}{{Lee} et~al.}{2014}]{Lee2014}
{Lee} E.~J.,  {Chiang} E.,   {Ormel} C.~W.,  2014, \mn@doi [\apj] {10.1088/0004-637X/797/2/95}, \href {http://adsabs.harvard.edu/abs/2014ApJ...797...95L} {797, 95}

\bibitem[\protect\citeauthoryear{{Lee}, {Karalis}  \& {Thorngren}}{{Lee} et~al.}{2022}]{Lee2022}
{Lee} E.~J.,  {Karalis} A.,   {Thorngren} D.~P.,  2022, arXiv e-prints, \href {https://ui.adsabs.harvard.edu/abs/2022arXiv220109898L} {p. arXiv:2201.09898}

\bibitem[\protect\citeauthoryear{{Lopez}}{{Lopez}}{2017}]{Lopez2017}
{Lopez} E.~D.,  2017, \mn@doi [\mnras] {10.1093/mnras/stx1558}, \href {https://ui.adsabs.harvard.edu/abs/2017MNRAS.472..245L} {472, 245}

\bibitem[\protect\citeauthoryear{{Lopez} \& {Fortney}}{{Lopez} \& {Fortney}}{2013}]{LopezFortney2013}
{Lopez} E.~D.,  {Fortney} J.~J.,  2013, \mn@doi [\apj] {10.1088/0004-637X/776/1/2}, \href {https://ui.adsabs.harvard.edu/abs/2013ApJ...776....2L} {776, 2}

\bibitem[\protect\citeauthoryear{{Lopez} \& {Fortney}}{{Lopez} \& {Fortney}}{2014}]{Lopez2014}
{Lopez} E.~D.,  {Fortney} J.~J.,  2014, \mn@doi [\apj] {10.1088/0004-637X/792/1/1}, \href {http://adsabs.harvard.edu/abs/2014ApJ...792....1L} {792, 1}

\bibitem[\protect\citeauthoryear{{Luo}, {Dorn}  \& {Deng}}{{Luo} et~al.}{2024}]{Luo2024}
{Luo} H.,  {Dorn} C.,   {Deng} J.,  2024, \mn@doi [arXiv e-prints] {10.48550/arXiv.2401.16394}, \href {https://ui.adsabs.harvard.edu/abs/2024arXiv240116394L} {p. arXiv:2401.16394}

\bibitem[\protect\citeauthoryear{{Luque} \& {Pall{\'e}}}{{Luque} \& {Pall{\'e}}}{2022}]{Luque2022}
{Luque} R.,  {Pall{\'e}} E.,  2022, \mn@doi [Science] {10.1126/science.abl7164}, \href {https://ui.adsabs.harvard.edu/abs/2022Sci...377.1211L} {377, 1211}

\bibitem[\protect\citeauthoryear{{Madhusudhan}, {Sarkar}, {Constantinou}, {Holmberg}, {Piette}  \& {Moses}}{{Madhusudhan} et~al.}{2023}]{Madhusudhan2023}
{Madhusudhan} N.,  {Sarkar} S.,  {Constantinou} S.,  {Holmberg} M.,  {Piette} A. A.~A.,   {Moses} J.~I.,  2023, \mn@doi [\apjl] {10.3847/2041-8213/acf577}, \href {https://ui.adsabs.harvard.edu/abs/2023ApJ...956L..13M} {956, L13}

\bibitem[\protect\citeauthoryear{{Misener} \& {Schlichting}}{{Misener} \& {Schlichting}}{2021}]{Misener2021}
{Misener} W.,  {Schlichting} H.~E.,  2021, \mn@doi [\mnras] {10.1093/mnras/stab895}, \href {https://ui.adsabs.harvard.edu/abs/2021MNRAS.503.5658M} {503, 5658}

\bibitem[\protect\citeauthoryear{{Misener} \& {Schlichting}}{{Misener} \& {Schlichting}}{2022}]{Misener2022}
{Misener} W.,  {Schlichting} H.~E.,  2022, \mn@doi [\mnras] {10.1093/mnras/stac1732}, \href {https://ui.adsabs.harvard.edu/abs/2022MNRAS.514.6025M} {514, 6025}

\bibitem[\protect\citeauthoryear{{Misener}, {Schlichting}  \& {Young}}{{Misener} et~al.}{2023}]{Misener2023}
{Misener} W.,  {Schlichting} H.~E.,   {Young} E.~D.,  2023, \mn@doi [\mnras] {10.1093/mnras/stad1910}, \href {https://ui.adsabs.harvard.edu/abs/2023MNRAS.524..981M} {524, 981}

\bibitem[\protect\citeauthoryear{{Misener}, {Schulik}, {Schlichting}  \& {Owen}}{{Misener} et~al.}{2024}]{Misener2024}
{Misener} W.,  {Schulik} M.,  {Schlichting} H.~E.,   {Owen} J.~E.,  2024, \mn@doi [arXiv e-prints] {10.48550/arXiv.2405.15221}, \href {https://ui.adsabs.harvard.edu/abs/2024arXiv240515221M} {p. arXiv:2405.15221}

\bibitem[\protect\citeauthoryear{{Mordasini}, {Alibert}  \& {Benz}}{{Mordasini} et~al.}{2009}]{Mordasini2009}
{Mordasini} C.,  {Alibert} Y.,   {Benz} W.,  2009, \mn@doi [\aap] {10.1051/0004-6361/200810301}, \href {https://ui.adsabs.harvard.edu/abs/2009A&A...501.1139M} {501, 1139}

\bibitem[\protect\citeauthoryear{{Mousis}, {Deleuil}, {Aguichine}, {Marcq}, {Naar}, {Aguirre}, {Brugger}  \& {Gon{\c{c}}alves}}{{Mousis} et~al.}{2020}]{Mousis2020}
{Mousis} O.,  {Deleuil} M.,  {Aguichine} A.,  {Marcq} E.,  {Naar} J.,  {Aguirre} L.~A.,  {Brugger} B.,   {Gon{\c{c}}alves} T.,  2020, \mn@doi [\apjl] {10.3847/2041-8213/ab9530}, \href {https://ui.adsabs.harvard.edu/abs/2020ApJ...896L..22M} {896, L22}

\bibitem[\protect\citeauthoryear{{Owen}}{{Owen}}{2020}]{Owen2020}
{Owen} J.~E.,  2020, \mn@doi [\mnras] {10.1093/mnras/staa2784}, \href {https://ui.adsabs.harvard.edu/abs/2020MNRAS.498.5030O} {498, 5030}

\bibitem[\protect\citeauthoryear{{Owen} \& {Jackson}}{{Owen} \& {Jackson}}{2012}]{OwenJackson2012}
{Owen} J.~E.,  {Jackson} A.~P.,  2012, \mn@doi [\mnras] {10.1111/j.1365-2966.2012.21481.x}, \href {https://ui.adsabs.harvard.edu/abs/2012MNRAS.425.2931O} {425, 2931}

\bibitem[\protect\citeauthoryear{{Owen} \& {Schlichting}}{{Owen} \& {Schlichting}}{2024}]{Owen2024}
{Owen} J.~E.,  {Schlichting} H.~E.,  2024, \mn@doi [\mnras] {10.1093/mnras/stad3972}, \href {https://ui.adsabs.harvard.edu/abs/2024MNRAS.528.1615O} {528, 1615}

\bibitem[\protect\citeauthoryear{{Owen} \& {Wu}}{{Owen} \& {Wu}}{2013}]{Owen2013}
{Owen} J.~E.,  {Wu} Y.,  2013, \mn@doi [\apj] {10.1088/0004-637X/775/2/105}, \href {https://ui.adsabs.harvard.edu/abs/2013ApJ...775..105O} {775, 105}

\bibitem[\protect\citeauthoryear{{Owen} \& {Wu}}{{Owen} \& {Wu}}{2016}]{Owen2016}
{Owen} J.~E.,  {Wu} Y.,  2016, \mn@doi [\apj] {10.3847/0004-637X/817/2/107}, \href {https://ui.adsabs.harvard.edu/abs/2016ApJ...817..107O} {817, 107}

\bibitem[\protect\citeauthoryear{{Owen} \& {Wu}}{{Owen} \& {Wu}}{2017}]{Owen2017}
{Owen} J.~E.,  {Wu} Y.,  2017, \mn@doi [\apj] {10.3847/1538-4357/aa890a}, \href {https://ui.adsabs.harvard.edu/abs/2017ApJ...847...29O} {847, 29}

\bibitem[\protect\citeauthoryear{{Petigura}, {Howard}  \& {Marcy}}{{Petigura} et~al.}{2013}]{Petigura2013}
{Petigura} E.~A.,  {Howard} A.~W.,   {Marcy} G.~W.,  2013, \mn@doi [Proceedings of the National Academy of Science] {10.1073/pnas.1319909110}, \href {https://ui.adsabs.harvard.edu/abs/2013PNAS..11019273P} {110, 19273}

\bibitem[\protect\citeauthoryear{{Petigura} et~al.,}{{Petigura} et~al.}{2018}]{Petigura2018}
{Petigura} E.~A.,  et~al., 2018, \mn@doi [\aj] {10.3847/1538-3881/aaa54c}, \href {https://ui.adsabs.harvard.edu/abs/2018AJ....155...89P} {155, 89}

\bibitem[\protect\citeauthoryear{{Petigura} et~al.,}{{Petigura} et~al.}{2022}]{Petigura2022}
{Petigura} E.~A.,  et~al., 2022, \mn@doi [\aj] {10.3847/1538-3881/ac51e3}, \href {https://ui.adsabs.harvard.edu/abs/2022AJ....163..179P} {163, 179}

\bibitem[\protect\citeauthoryear{{Piaulet-Ghorayeb} et~al.,}{{Piaulet-Ghorayeb} et~al.}{2024}]{Piaulet-Ghorayeb2024}
{Piaulet-Ghorayeb} C.,  et~al., 2024, \mn@doi [\apjl] {10.3847/2041-8213/ad6f00}, \href {https://ui.adsabs.harvard.edu/abs/2024ApJ...974L..10P} {974, L10}

\bibitem[\protect\citeauthoryear{{Pierrehumbert}}{{Pierrehumbert}}{2023}]{Pierrehumbert2023}
{Pierrehumbert} R.~T.,  2023, \mn@doi [\apj] {10.3847/1538-4357/acafdf}, \href {https://ui.adsabs.harvard.edu/abs/2023ApJ...944...20P} {944, 20}

\bibitem[\protect\citeauthoryear{{Raymond}, {Boulet}, {Izidoro}, {Esteves}  \& {Bitsch}}{{Raymond} et~al.}{2018}]{Raymond2018}
{Raymond} S.~N.,  {Boulet} T.,  {Izidoro} A.,  {Esteves} L.,   {Bitsch} B.,  2018, \mn@doi [\mnras] {10.1093/mnrasl/sly100}, \href {https://ui.adsabs.harvard.edu/abs/2018MNRAS.479L..81R} {479, L81}

\bibitem[\protect\citeauthoryear{{Rivera} et~al.,}{{Rivera} et~al.}{2005}]{Rivera2005}
{Rivera} E.~J.,  et~al., 2005, \mn@doi [\apj] {10.1086/491669}, \href {https://ui.adsabs.harvard.edu/abs/2005ApJ...634..625R} {634, 625}

\bibitem[\protect\citeauthoryear{{Rogers}}{{Rogers}}{2015}]{Rogers2015}
{Rogers} L.~A.,  2015, \mn@doi [\apj] {10.1088/0004-637X/801/1/41}, \href {https://ui.adsabs.harvard.edu/abs/2015ApJ...801...41R} {801, 41}

\bibitem[\protect\citeauthoryear{{Rogers} \& {Owen}}{{Rogers} \& {Owen}}{2021}]{Rogers2021}
{Rogers} J.~G.,  {Owen} J.~E.,  2021, \mn@doi [\mnras] {10.1093/mnras/stab529}, \href {https://ui.adsabs.harvard.edu/abs/2021MNRAS.503.1526R} {503, 1526}

\bibitem[\protect\citeauthoryear{{Rogers}, {Gupta}, {Owen}  \& {Schlichting}}{{Rogers} et~al.}{2021}]{Rogers2021b}
{Rogers} J.~G.,  {Gupta} A.,  {Owen} J.~E.,   {Schlichting} H.~E.,  2021, \mn@doi [\mnras] {10.1093/mnras/stab2897}, \href {https://ui.adsabs.harvard.edu/abs/2021MNRAS.508.5886R} {508, 5886}

\bibitem[\protect\citeauthoryear{{Rogers}, {Jan{\'o} Mu{\~n}oz}, {Owen}  \& {Makinen}}{{Rogers} et~al.}{2023a}]{Rogers2023}
{Rogers} J.~G.,  {Jan{\'o} Mu{\~n}oz} C.,  {Owen} J.~E.,   {Makinen} T.~L.,  2023a, \mn@doi [\mnras] {10.1093/mnras/stad089}, \href {https://ui.adsabs.harvard.edu/abs/2023MNRAS.519.6028R} {519, 6028}

\bibitem[\protect\citeauthoryear{{Rogers}, {Schlichting}  \& {Owen}}{{Rogers} et~al.}{2023b}]{Rogers2023b}
{Rogers} J.~G.,  {Schlichting} H.~E.,   {Owen} J.~E.,  2023b, \mn@doi [\apjl] {10.3847/2041-8213/acc86f}, \href {https://ui.adsabs.harvard.edu/abs/2023ApJ...947L..19R} {947, L19}

\bibitem[\protect\citeauthoryear{{Rogers}, {Owen}  \& {Schlichting}}{{Rogers} et~al.}{2024a}]{Rogers2024a}
{Rogers} J.~G.,  {Owen} J.~E.,   {Schlichting} H.~E.,  2024a, \mn@doi [\mnras] {10.1093/mnras/stae563}, \href {https://ui.adsabs.harvard.edu/abs/2024MNRAS.529.2716R} {529, 2716}

\bibitem[\protect\citeauthoryear{{Rogers}, {Schlichting}  \& {Young}}{{Rogers} et~al.}{2024b}]{Rogers2024b}
{Rogers} J.~G.,  {Schlichting} H.~E.,   {Young} E.~D.,  2024b, \mn@doi [\apj] {10.3847/1538-4357/ad5287}, \href {https://ui.adsabs.harvard.edu/abs/2024ApJ...970...47R} {970, 47}

\bibitem[\protect\citeauthoryear{{Rogers}, {Dorn}, {Aditya Raj}, {Schlichting}  \& {Young}}{{Rogers} et~al.}{2025}]{Rogers2025}
{Rogers} J.~G.,  {Dorn} C.,  {Aditya Raj} V.,  {Schlichting} H.~E.,   {Young} E.~D.,  2025, \mn@doi [\apj] {10.3847/1538-4357/ad9f61}, \href {https://ui.adsabs.harvard.edu/abs/2025ApJ...979...79R} {979, 79}

\bibitem[\protect\citeauthoryear{{Santos} et~al.,}{{Santos} et~al.}{2004}]{Santos2004}
{Santos} N.~C.,  et~al., 2004, \mn@doi [\aap] {10.1051/0004-6361:200400076}, \href {https://ui.adsabs.harvard.edu/abs/2004A&A...426L..19S} {426, L19}

\bibitem[\protect\citeauthoryear{{Schlichting} \& {Young}}{{Schlichting} \& {Young}}{2022}]{Schlichting2022}
{Schlichting} H.~E.,  {Young} E.~D.,  2022, \mn@doi [PSJ] {10.3847/PSJ/ac68e6}, \href {https://ui.adsabs.harvard.edu/abs/2022PSJ.....3..127S} {3, 127}

\bibitem[\protect\citeauthoryear{{Schulik} \& {Booth}}{{Schulik} \& {Booth}}{2022}]{Schulik2022}
{Schulik} M.,  {Booth} R.,  2022, arXiv e-prints, \href {https://ui.adsabs.harvard.edu/abs/2022arXiv220707144S} {p. arXiv:2207.07144}

\bibitem[\protect\citeauthoryear{Szurgot}{Szurgot}{2015}]{Szurgot2015}
Szurgot M.,  2015, in 46th Lunar and Planetary Science Conference.

\bibitem[\protect\citeauthoryear{{Tang}, {Fortney}  \& {Murray-Clay}}{{Tang} et~al.}{2024}]{Tang2024}
{Tang} Y.,  {Fortney} J.~J.,   {Murray-Clay} R.,  2024, \mn@doi [arXiv e-prints] {10.48550/arXiv.2410.08577}, \href {https://ui.adsabs.harvard.edu/abs/2024arXiv241008577T} {p. arXiv:2410.08577}

\bibitem[\protect\citeauthoryear{{Thao} et~al.,}{{Thao} et~al.}{2024}]{Thao2024}
{Thao} P.~C.,  et~al., 2024, \mn@doi [\aj] {10.3847/1538-3881/ad81d7}, \href {https://ui.adsabs.harvard.edu/abs/2024AJ....168..297T} {168, 297}

\bibitem[\protect\citeauthoryear{{Thorngren} \& {Fortney}}{{Thorngren} \& {Fortney}}{2019}]{Thorngren2019}
{Thorngren} D.,  {Fortney} J.~J.,  2019, \mn@doi [\apjl] {10.3847/2041-8213/ab1137}, \href {https://ui.adsabs.harvard.edu/abs/2019ApJ...874L..31T} {874, L31}

\bibitem[\protect\citeauthoryear{{Vach} et~al.,}{{Vach} et~al.}{2024}]{Vach2024a}
{Vach} S.,  et~al., 2024, \mn@doi [\aj] {10.3847/1538-3881/ad3108}, \href {https://ui.adsabs.harvard.edu/abs/2024AJ....167..210V} {167, 210}

\bibitem[\protect\citeauthoryear{{Van Eylen}, {Agentoft}, {Lundkvist}, {Kjeldsen}, {Owen}, {Fulton}, {Petigura}  \& {Snellen}}{{Van Eylen} et~al.}{2018}]{VanEylen2018}
{Van Eylen} V.,  {Agentoft} C.,  {Lundkvist} M.~S.,  {Kjeldsen} H.,  {Owen} J.~E.,  {Fulton} B.~J.,  {Petigura} E.,   {Snellen} I.,  2018, \mn@doi [\mnras] {10.1093/mnras/sty1783}, \href {https://ui.adsabs.harvard.edu/abs/2018MNRAS.479.4786V} {479, 4786}

\bibitem[\protect\citeauthoryear{{Vazan}, {Sari}  \& {Kessel}}{{Vazan} et~al.}{2022}]{Vazan2022}
{Vazan} A.,  {Sari} R.,   {Kessel} R.,  2022, \mn@doi [\apj] {10.3847/1538-4357/ac458c}, \href {https://ui.adsabs.harvard.edu/abs/2022ApJ...926..150V} {926, 150}

\bibitem[\protect\citeauthoryear{{Venturini}, {Alibert}  \& {Benz}}{{Venturini} et~al.}{2016}]{Venturini2016}
{Venturini} J.,  {Alibert} Y.,   {Benz} W.,  2016, \mn@doi [\aap] {10.1051/0004-6361/201628828}, \href {https://ui.adsabs.harvard.edu/abs/2016A&A...596A..90V} {596, A90}

\bibitem[\protect\citeauthoryear{{Weiss} \& {Marcy}}{{Weiss} \& {Marcy}}{2014}]{Weiss2014}
{Weiss} L.~M.,  {Marcy} G.~W.,  2014, \mn@doi [\apj] {10.1088/2041-8205/783/1/L6}, \href {https://ui.adsabs.harvard.edu/abs/2014ApJ...783L...6W} {783, L6}

\bibitem[\protect\citeauthoryear{{Wolfgang}, {Rogers}  \& {Ford}}{{Wolfgang} et~al.}{2016}]{Wolfgang2016}
{Wolfgang} A.,  {Rogers} L.~A.,   {Ford} E.~B.,  2016, \mn@doi [\apj] {10.3847/0004-637X/825/1/19}, \href {https://ui.adsabs.harvard.edu/abs/2016ApJ...825...19W} {825, 19}

\bibitem[\protect\citeauthoryear{{Wu} \& {Lithwick}}{{Wu} \& {Lithwick}}{2013}]{Wu_Lithwick2013}
{Wu} Y.,  {Lithwick} Y.,  2013, \mn@doi [\apj] {10.1088/0004-637X/772/1/74}, \href {https://ui.adsabs.harvard.edu/abs/2013ApJ...772...74W} {772, 74}

\bibitem[\protect\citeauthoryear{{Young}, {Stixrude}, {Rogers}, {Schlichting}  \& {Marcum}}{{Young} et~al.}{2024}]{Young2024}
{Young} E.~D.,  {Stixrude} L.,  {Rogers} J.~G.,  {Schlichting} H.~E.,   {Marcum} S.~P.,  2024, \mn@doi [arXiv e-prints] {10.48550/arXiv.2408.11321}, \href {https://ui.adsabs.harvard.edu/abs/2024arXiv240811321Y} {p. arXiv:2408.11321}

\bibitem[\protect\citeauthoryear{{Zeng} et~al.,}{{Zeng} et~al.}{2019}]{Zeng2019}
{Zeng} L.,  et~al., 2019, \mn@doi [Proceedings of the National Academy of Science] {10.1073/pnas.1812905116}, \href {https://ui.adsabs.harvard.edu/abs/2019PNAS..116.9723Z} {116, 9723}

\makeatother
\end{thebibliography}

%\begin{thebibliography}{99}
%\end{thebibliography}

%%%%%%%%%%%%%%%%%%%%%%%%%%%%%%%%%%%%%%%%%%%%%%%%%%

%%%%%%%%%%%%%%%%% APPENDICES %%%%%%%%%%%%%%%%%%%%%

% \appendix

% \section{Some extra material}

% If you want to present additional material which would interrupt the flow of the main paper,
% it can be placed in an Appendix which appears after the list of references.

%%%%%%%%%%%%%%%%%%%%%%%%%%%%%%%%%%%%%%%%%%%%%%%%%%

% Don't change these lines
\bsp	% typesetting comment
\label{lastpage}
\end{document}